\documentclass{emulateapj}
\usepackage{graphicx}
\usepackage{subfigure}
\usepackage{natbib}

\shorttitle{Metallicities and Kinematics in WLM}
\shortauthors{Leaman et al.}

\begin{document}

\title{Stellar Metallicities and Kinematics in a Gas-Rich Dwarf Galaxy: First Calcium Triplet Spectroscopy of RGB Stars in WLM}
\author{Ryan Leaman$^{1}$, Andrew A. Cole$^{2}$, Kim A. Venn$^{1}$, Eline Tolstoy$^{3}$, Mike J. Irwin$^{4}$,\\ Thomas Szeifert$^{5}$, Evan D. Skillman$^{6}$, Alan W. McConnachie$^{1,7}$}
\affil{$^{1}$University of Victoria, Canada, $^{2}$University of Tasmania, Australia, $^{3}$Kapteyn Institute, The Netherlands, $^{4}$Institute of Astronomy, U.K., $^{5}$ESO-Chile, Chile, $^{6}$University of Minnesota, U.S., $^{7}$National Research Council of Canada, Herzberg Institute of Astrophysics, Canada}
\email{rleaman@uvic.ca}

\begin{abstract}
We present the first determination of the radial velocities and 
metallicities of 78 red giant stars in the isolated dwarf irregular 
galaxy WLM.  Observations of the calcium II triplet  in these stars
were made with FORS2 at the VLT-UT2 in two separated fields of view in WLM, and the [Fe/H] values were conformed to the \cite{CG97} ([Fe/H]$_{CG97}$) metallicity scale. The mean metallicity is $\langle$[Fe/H]$\rangle = -1.27 \pm 0.04$ dex, with a standard deviation of $\sigma = 0.37$.   We find that the stars in the inner field are more metal rich by $\Delta$[Fe/H]$=0.30 \pm 0.06$ dex.   
These results are in agreement with previous photometric studies that found 
a radial population gradient, as well as the expectation
of higher metallicities in the central star forming regions.
Age estimates using Victoria-Regina stellar models show that 
the youngest stars in the sample ($< 6$ Gyr) are 
more metal rich by $\Delta$[Fe/H]$= 0.32 \pm 0.08$ dex.    These stars
also show a lower velocity dispersion at all elliptical radii compared
to the metal-poor stars.   Kinematics for the whole red giant sample suggest
a velocity gradient approximately half that of the gas rotation curve, with the stellar component occupying a thicker disk decoupled from the HI rotation plane.
Taken together, the kinematics, metallicities, and ages in our sample
suggest a young metal-rich, and kinematically cold stellar population 
in the central gas-rich regions of WLM, surrounded by a separate dynamically 
hot halo of older, metal poor stars.   
\end{abstract}


\section{Introduction}
Dwarf galaxies play a critical role in our understanding of the assembly 
of galaxies in LCDM cosmologies.  With masses of $10^{8}$ to $10^{9}$ 
M$_{\odot}$, these galaxies are thought to be similar to the proto-galactic
fragments that merged and collapsed to form large galaxies 
(e.g., \citealt{NFW97, Moore99, Madau01}).  Analysing the survival of 
these low mass objects, particularly through reionisation 
\citep{Ricotti05, Gnedin06}, is crucial to constraining galaxy formation
models.  For example, what was the minimum halo mass that could retain its baryons through reionisation?  Theoretical constraints on low mass galaxies are also provided by examining the 
detailed characteristics of the various types of dwarf galaxies 
(irregulars, spheroidals, and the new ultra faint dwarfs) and the 
connections between them.  Are dwarf irregular galaxies simply dSph which have undergone recent gas mergers \citep{Demers06, Brook07}?  Are transition galaxies gas-rich dwarf galaxies being subject to ram pressure stripping (e.g., Pegasus dwarf galaxy; \citealt{Alan07})?  Are signatures of thick stellar disks or spheroidal components expected in dwarf irregular galaxies?  How will the dynamics of the stellar populations compare to the gas motions in the low mass galaxies?  
This line of research can be carried out 
on the nearby Local Group galaxies, thus defining near-field cosmology.

Dwarf irregular (dIrr) galaxies hold a special status in the analysis of the
Local Group galaxies because most are relatively isolated.   Detailed
studies of the nearby dwarf spheroidal galaxies have revealed complex
and varied star formation histories that have left behind distinct
stellar populations \citep{Tolstoy04, Battaglia06, Bosler07}.  
However, interpretation of the kinematics of these stellar populations, 
and therefore evolution of the nearby dSph galaxies, is complicated by the 
fact that they exist in the dark matter halos of the MW and M31 
(i.e., their stellar populations have likely been tidally stirred).  
On the contrary, dwarf irregular galaxies are relatively \emph{isolated} 
low mass galaxies.  Evolved stellar populations in dIrrs may prove to 
be excellent tracers of the dynamical history of low mass dwarf galaxies 
at early times, and therefore excellent comparisons for galaxy formation 
models.  

However, there have been few studies of the kinematics of the stellar populations 
in isolated dIrrs and no detailed spectroscopic analysis of their $\emph{older}$ stellar populations due to their distances.\footnote{We note that \cite{Tolstoy01} examined 
the calcium II triplet feature in 23 RGB stars in the closest dIrr, NGC 6822 
($V_{TRGB}\sim 21$), and found most stars were young and metal-rich, 
representative of the dominant young population at the brightest magnitudes.}
This is significant because various galaxy formation scenarios predict 
different characteristics for the stellar populations in early dwarf 
galaxies; e.g., simulations by \cite{Mayer06} predict disk like systems
that become more spheroidal through tidal interactions and ram pressure
stripping, whereas \cite{Kaufmann07} suggest that dwarf galaxies start 
out as thick, puffy systems and through gas losses and tidal interactions
become more disk like.

In this paper, we present the first spectroscopic analysis 
of the calcium II triplet (CaT) feature 
in a sample of RGB stars in the dIrr galaxy WLM.
WLM is a typical low-luminosity, high gas fraction late-type dwarf 
irregular galaxy.  A summary of its fundamental parameters is
listed in Table 1.  The nearest neighbour to WLM is the Cetus dSph, which lies 200 kpc away \citep{Whiting99}. 
WLM's distance to the Milky Way is estimated at $\sim 932 \pm 33$ kpc \citep{Alan05}, 
and its separation from M31 is $\sim$820 kpc, therefore 
this is one of the most isolated galaxies in the Local Group.   
WLM has a heliocentric velocity of -130 km s$^{-1}$ \citep{Jackson04} and a modest velocity with respect to the Local Group barycentre, $-29$ km s$^{-1}$, implying that it may have recently passed apocentre and is turning around.  Additionally, WLM lies out of the Galactic plane, which minimizes 
foreground contamination and reddening.  

High dispersion spectra have been taken for a few bright A and B-type 
supergiant stars in WLM \citep{Venn03, Bresolin06, Urbaneja08}.  Detailed
analyses of these stars provide the present day metallicities and abundance ratios ([Fe/H]$=-0.38 \pm 0.2$, \citealt{Venn03}; [O/H]$\sim -0.85$, \citealt{Bresolin06}; [Z]$=-0.87 \pm 0.06$, \citealt{Urbaneja08}), but offer little information on the intermediate or old age 
populations.  The older red giant branch stars in these isolated dIrrs 
are too faint for high dispersion spectroscopic analyses, even with 8m 
class telescopes.
Studies of HII regions from emission line spectroscopy yield [O/H] = $-0.83$
\citep{Skillman89, Hodge95, Lee05}, but provide no information about the 
chemistry of the gas in the early stages of the galaxy.  
Interestingly, WLM has revealed minor discrepancies in chemistry between 
the young stars and H II regions - possibly due to inhomogeneous mixing (e.g., \citealt{Venn03} found [O/H] = $-0.21$ in one A supergiant; \citealt{Lee05} find [O/H] = $-0.83 \pm 0.06$ from emission line measurements in two HII regions, which included the [OIII] $\lambda$ 4363 line).
Neutral gas studies in WLM have been used to map the HI envelope extent
and small scale spatial and velocity structures
\citep{Huchtmeier81, Barnes04, Jackson04, Kepley07}. 

The old population in WLM was first sampled by \cite{Hodge99}
who derived a metallicity from isochrone fitting to deep HST imaging
of the lone globular cluster, WLM-1.  
HST photometry and wide field INT imaging 
\citep{Minniti97, Rejkuba00, Alan05} also identified young and
old stellar populations in the form of an extended blue main 
sequence, and a horizontal branch on the CMD.  These photometric
surveys were also used to find the distance and reddening to WLM, 
and to estimate the range in metallicity on the RGB.
Photometric analyses of C and M stars in WLM by \cite{Battinelli04} 
have argued against the presence of an old extended halo in WLM, 
opposite to the conclusion from \cite{Minniti97} based on 
the interpretation of their CMDs.  
However, differential reddening within WLM may be affecting all of these
photometric analyses; the recent Spitzer IRAC survey of AGB stars 
in WLM \citep{Jackson07a} has shown the patchy presence of dust 
throughout WLM.   

The use of the empirically calibrated near infrared calcium triplet lines provides a new method for studying the stellar population
in WLM.  Situated at $\lambda \sim$8498, 8542, 8662 ${\rm \AA}$, they are 
optimally located with minimal contamination from other spectral features 
and near the peak in flux for these evolved red stars.  
The summed equivalent widths (EWs) of these lines are well calibrated to allow a representative placement of a star onto a given [Fe/H]\footnote{The notation [Fe/H] = log(Fe/H)$_{*}$ - log(Fe/H)$_{\sun}$} scale, and sensitive enough from medium resolution spectra to perform well out to large distances in the local volume.  The metallicity index is also well correlated with the iron abundances ([Fe/H]) determined by \cite{CG97} from high dispersion spectroscopy of Galactic globular clusters.  This scale shows a linear correlation with the CaT W' index, unlike the Zinn \& West \citep{ZW84} scale based upon the Q$_{39}$ spectrophotometric index.
Previous large scale calibration studies \citep{Rutledge97, Cole04, Carrera07b, Battaglia07} confirmed the robustness of 
the CaT method over a range of ages ($0.25\leq$ Gyr $\leq 13$) and 
over the metallicity range expected for the stars in a gas rich 
dIrr galaxy ($-2.5\lesssim$ [Fe/H] $\lesssim +0.47$).  A growing 
number of CaT studies have been carried out for several Local Group 
galaxies, including the Magellanic clouds \citep{Pont04, Cole05, 
Grocholski06, Carrera07} and dSph galaxies 
\citep{Tolstoy04, Battaglia06, Koch06, Bosler07}, which further 
tests the robustness of the CaT method in different environments.

In the following sections, we discuss the observations, the 
data reduction methods, and spectral analysis of CaT spectroscopy
adopted in this paper.   These sections are
particularly important since this work represents a CaT analysis of
some of the faintest RGB stars for which velocities \textit{and} metallicities
have been determined at moderate S/N.  The main challenge has been 
to minimize the 
velocity and metallicity errors.  Although the final uncertainties are 
slightly larger than CaT surveys of closer galaxies, we have been 
able to determine a new metallicity distribution function for WLM 
and characterize the spatial and velocity variations in its stellar 
populations for the first time. 

\section{Observations and Data Reductions}
\subsection{\textit{Target Selection and Photometric Calibration}}
Relatively isolated and bright stars near the tip of the RGB (V$_{mag}\sim 23$) were selected from 150 second V and I band preimaging exposures from FORS2 on the VLT \citep{FORS2}.  The data were bias corrected and flatfielded in IRAF\footnote{IRAF(Image Reduction and Analysis Facility) is distributed by the National Optical Astronomy Observatory, which is operated by the Association of Universities for Research in Astronomy, Inc., under cooperative agreement with the National Science Foundation} and instrumental photometry was obtained using DAOPHOT/ALLFRAME \citep{Stetson94}.  Stars were selected from the tip of the RGB (TRGB) to those within half a magnitude below the RGB tip (instrumental I band), and colors consistent with RGB membership.  This meant that stars were selected to have instrumental colours spanning the apparent width of the RGB (i.e., constrained to a colour range of 1.1 magnitudes) to avoid heavy contamination by red supergiants and M stars.  Our targets were selected to encompass a broad area of the galaxy's high density (gas and stellar) regions, as well as lower density outer areas.  

The FORS2 preimaging photometry was then matched to extant INT WFC V, i band photometry of a 0.25 square degree region centered on WLM (see \cite{Alan05} for further details), due to the preimaging being taken under non-photometric conditions.  This allowed us to use standard magnitudes and colours for our analysis.  A list of INT WFC stars for matching was created based on colour, magnitude, and location, and these stars were further filtered based on sharpness and ellipse parameters.   As the stars were uniformly selected to be bright and isolated,  we acknowledge that a selection effect may appear in areas of low stellar density.  However, this is unavoidable in resolved stellar spectroscopy at such large distances, and simply necessitates that care be taken in the interpretation of the results.  Four globular clusters (47Tuc, NGC1851, NGC1904, M15) spanning a range of metallicities, were also chosen for calibration purposes, as described in $\S$3.1.1. 

\subsection{\textit{Data Acquisition and Reduction}}
 The observations for this study of WLM and the four calibrating clusters were taken during several nights at the VLT in late 2003 (see Table 1).  The MXU (Mask eXchange Unit) mode was used with the FORS2 instrument at the Cassegrain unit of VLT's UT4 (Yepun) telescope, note this was before the instrument was relocated to UT1 (Antu) in 2004.  The 83 RGB stellar targets ranged in magnitude from 22.1 $\leq$ V$_{mag}\leq$ 24.0, requiring exposure times on the order of 40 minutes for individual images even with an 8m class telescope. Slit acquisition images were taken for approximately 150 seconds in Bessell I band, to confirm the pointing and slit alignment prior to science exposures. The parameters for the object science exposures on the observational dates in question are shown in Table 2.

Figure \ref{fig:fov} shows a representative view of the FORS2/MXU fields used in this study of WLM.  For reference, the INT WFC fields are the four red rectangles.  The FORS2 instrument was used with the MXU in order to provide selectable custom cut slit plates; a configuration which allowed the preselected RGB stars to be fit within $1^{''}\times8^{''}$ slits to minimize the required observation time spent for the program.  A spectroscopic order separation filter (OG590+32) and the standard resolution collimator were used in conjunction with a volume phased holographic grism to obtain the stellar spectra for each slit target.  The grism, designated GRIS$\_$1028+29, provided wavelength coverage from roughly 7730${\rm \AA}$ to 9480${\rm \AA}$ with a central wavelength of $\lambda_{c}=8600{\rm \AA}$, and a dispersion of 28.3${\rm \AA}$/mm.  FORS2 is equipped with a mosaic of two red-optimized 2k$\times$4k MIT CCDs ($15\mu$m pixels) with very low fringe amplitude in the spectral range of the CaT lines.  With 2$\times$2 binning and 100khz readout characteristics, and the above mentioned optical path, the effective field of view across the instrument was $6.8^{'}\times5.7^{'}$.  
The two component chips have a pixel scale of 0.252$^{''}$/pixel, where the pixel size is 0.86${\rm \AA}$ ($2 \times 2$ binning).   This setup produced observations with 
resolving power R~$\sim 3400$.  The gain for both chips was 
0.7 ADUs$/e^{-}$, with the readnoise being 2.7$e^{-}$ and 3.15$e^{-}$, 
respectively.  Each of the large blue boxes in Figure \ref{fig:fov} is 
representative of the full two chip CCD field of view.  The representative seeing conditions for the observations at VLT ranged from $0.61^{''} \leq$ FWHM $\leq 1.52^{''}$ over the course of our science exposures.

Once obtained in service mode at the telescope, the data were reduced using a variety of standard IRAF tasks for bias, overscan and flatfield corrections to the two dimensional images.  Typically each science image used a bias correction combined from 10 individual bias exposures, along with 5 individual screen flat field exposures for each of the chips, which were reduced independently.  The objects were cleaned for cosmic rays using several iterations of the \textsc{cosmicrays} task. In order to minimize any error in the final aperture extraction or wavelength calibration, custom tasks were run to remove warping in the stellar trace and sky lines (particularly the far corners of the chips) due to light path distortions or chip alignment issues.  A custom IRAF script was used to provide a preliminary correction to the drooping artifact of the stellar trace as a function of row position on the CCD.  A second script was then used to linearize and orthogonalize the dispersion coordinate and spatial coordinate to optimize subtraction of the sky lines (see \citealt{Grocholski06}).  Following these corrections, aperture extraction was run on all science spectra, resulting in one dimensional spectral images for all the stars in the sample.  We used a standard sky line atlas in conjunction with the \textsc{identify} and \textsc{reidentify} tasks to provide dispersion solutions for the wavelength calibration of the spectra.  
Typical line fits involved $\sim40$ OH sky emission features taken from 
\cite{Osterbrock92}.  The typical accuracy of the wavelength solution is within $\sim 0.04 {\rm \AA}$ based on the rms of the solutions.  Finally, the task \textsc{continuum} removed the curvature of the spectrum and normalized the flux distribution to allow for the most accurate measurements of the regions of interest (the calcium triplet lines).  

Prior to combination, we adjusted the wavelength zeropoint of each spectrum 
for the heliocentric correction.  This was a non-trivial concern for the 
radial velocity measurements as our individual exposures were spread over 
several months (See Table 2).  A date dependent shift was applied to 
each spectrum, such that their observed velocities were corrected back to 
a common epoch at the date of the first observation.  
This resulted in at most a $\sim1.2{\rm \AA}$ correction to the spectra.
The final spectra include
the median of eight individual exposures for each of 79 RGB stars (four stars were thrown out due to spectral contamination or photometry matching issues).  
The typical signal to noise ratio for our combined images ranged 
between $16\lesssim$ (S/N) $\lesssim 23$ per pixel (1 pixel = 0.86${\rm \AA}$).

\section{Spectral Analysis}
\subsection{\textit{Equivalent Width Measurements}}
The choice of techniques for equivalent width measurements had to be carefully considered, given the moderate signal to noise of our combined spectra ($\sim 20$).  As noted by \cite{Cole04}, the line wings are typically underestimated with a pure Gaussian profile fit, and are much more accurately modeled with the sum of a Gaussian and Lorentzian fit.  
However, for low signal-to-noise spectra, when the line FWHM are on the order of the spectral resolution, the contaminating noise features effectively nullify any difference between using a Gaussian, Lorentzian, or the sum of the two. Therefore an accurate characterization of the spectrographic setup is necessary in low resolution Calcium triplet spectroscopic studies.
For this reason, the choice was made to use a simple pixel to pixel integration of the line profiles.  Due to the low signal-to-noise of the spectra, the integration was taken over the range where the line wings intersected the global continuum level. The continuum normalized spectra had an average continuum at unity already, no adjustment was made to the flux zeropoint.  The integration yields a wavelength, core density, and equivalent width for each line.  Multiple independent measurements were made by remeasuring the same line repeatedly, with a systematic deviation in equivalent width of approximately 1\%.  Figure \ref{fig:fourspec} shows a sample spectrum in the region of the Ca II triplet lines for one star in our sample.

As a consistency check, we performed simultaneous measurements on the reduced calibrating clusters (which had a much higher signal to noise) using both the pixel to pixel integration method and the profile fitting programs from \cite{Cole04}.  
The linear fit between the methods' measurements (see Figure \ref{fig:sigewcomp}) was used to develop a transformation function which put our equivalent widths onto the same scale as in \cite{Cole04}, thereby allowing us to adopt a variety of their calibrations as described in the next section.  

\subsubsection{Placement onto the Metallicity Scale}
\cite{ADC91} first showed the usefulness of the CaT feature as a metallicity indicator in individual evolved giant branch stars.  Using the summed equivalent widths of the CaT lines, as well as the star's V magnitude above the horizontal branch, a reduced equivalent width could be formulated.  When observed for Galactic globular cluster stars, which also had high dispersion spectroscopic [Fe/H] estimates, this low resolution CaT estimator could be used to provide an empirical metallicity index.  Several studies (e.g., \citealt{Rutledge97, Cole04}) have since explored these empirical calibrations over a large range of stellar ages and
metallicities. \cite{Cole04} found minimal deterioration in the CaT calibration over 
extremes in age and metallicity.  This is particularly relevant for dwarf galaxy
studies where we expect a mixed stellar population based on CMD analyses
(e.g., \citealt{Dolphin00}).  In this paper, we adopt the calibrations 
determined by \cite{Cole04} based on many clusters which were made 
with the same instrument (FORS2) at the VLT, and adopted similar 
data reduction techniques as used here.  
The spread in cluster ages in the full sample studied by 
\cite{Cole04} is ideal for analyzing a dIrr galaxy like WLM.  Recently, \cite{Battaglia07}, \cite{Carrera07}, and \cite{Battaglia08} have confirmed the robustness 
of the CaT spectroscopic method as a means for metallicity estimates 
in dwarf galaxies.  These studies 
specifically checked the appropriateness of the CaT index as a proxy 
for [Fe/H] in cases of varying [$\alpha$/Fe], [Ca/Fe],  and age.  
Additionally, studies by \cite{Rutledge97} and \cite{Cenarro01} showed that 
the CaT index transforms in a well understood way between different 
authors' studies. 

To use any of the calibrations requires a summed equivalent width determination for each star.  Each of the three calcium triplet line measurements were combined in an 
unweighted fashion to yield a summed equivalent width per star.
\begin{center}
\begin{equation}
\Sigma W = W_{8498}+W_{8542}+W_{8662}
\end{equation}
\end{center}
The justification to use all three lines will be discussed in $\S$3.2.2 and $\S$3.2.3 with respect to our errors.  With this relation it is now possible to form the calcium index, $W'$ defined as:
\begin{center}
\begin{equation}
W' = \Sigma W + \beta(V - V_{HB})    
\end{equation}
\end{center}
The term in the parentheses provides a correction for 
the changes in T$_{eff}$ and $log(g)$ for stars in different phases on 
the red giant branch.  
A cooler temperature and lower surface gravity play non-trivial roles 
in the formation of the CaT line profiles and the continuum in these
evolutionary stages.
Theoretical and empirical work \citep{Jorgenson92, Cenarro02} has confirmed this 
complicated interplay of the calcium line strengths with stellar 
parameters such as temperature and gravity.  Thus, this term is important 
in removing the gravity dependence of the lines with respect to the 
continuum in the CaT analysis.  Our V magnitudes are taken from the 
INT WFC catalogue and associated database as mentioned in $\S$2.1.  
We adopted the horizontal branch at V$_{HB}=25.71\pm0.09$ mag 
\citep{Rejkuba00}, and take $\beta=0.73\pm0.04 {\rm \AA}$ mag$^{-1}$ 
from \cite{Cole04}.  Using the \cite{CG97} scale, the calcium index 
is converted to a metallicity ([Fe/H]$_{CG97}$) as follows:
\begin{center}
\begin{equation}
[Fe/H]_{CG97} = (0.362\pm0.014)W' - (2.966\pm0.032)
\end{equation}
\end{center}

The zero point and slope were determined by \cite{Cole04}, 
and estimated as accurate to $\leq 4\%$.  
This means that the majority of our uncertainties are from other factors,
which we will discuss below in the following section.
\\

Figure \ref{fig:vvsigewfl} shows a plot of the summed calcium II 
equivalent widths ($\Sigma$W) against the V magnitude above the 
horizontal branch per star.  
The fiducial solid lines show the spacing in metallicity given by 
the calibration from \cite{Cole04}.

\subsubsection{Error Analysis}
To examine the various sources of uncertainty in our calculations, 
all errors from the equivalent width measurements to placement
on the [Fe/H]$_{CG97}$ scale were recorded.   As no standalone error estimate was provided by the pixel integration routine used in measuring a line equivalent width, $W$,  we adopted the Cayrel formula 
(see Eq. 7; \citealt{Cayrel88}, also \citealt{Battaglia07}) as:

\begin{center}
\begin{equation} 
\langle\Delta W_{n}^{2}\rangle ^{\frac{1}{2}} \simeq (1.6 (FWHM_{n} \delta x)^{\frac{1}{2}} \epsilon) + (0.10 W_{n})
\end{equation}
\end{center}
where the pixel size is $\delta x = 0.86{\rm \AA}$, and the average rms continuum accuracy is $\epsilon\sim\frac{1}{(S/N)_{avg}}$.  As the pixel integration measurements did not contain information on the FWHM of the line, that value was determined as follows.  A test set of Gaussian and Lorentzian fits were made to the strongest line (8542 ${\rm \AA}$) in 18 stars, using the line profile fitting routine \textsc{splot}.  
A Gaussian-Lorentzian blended estimate of the FWHM was estimated from  an arithmetic average of the two fit values.  These were then plotted as a function of  measured equivalent width (from the pixel integration method).  A linear regression was fit;  an estimate of the FWHM to be used in this formula for the rest of the stars was then determined based on their equivalent width values.  The FWHM values for the sub-sample of stars used in this step showed a standard deviation of $\sigma_{fwhm} = 1.06 {\rm \AA}$.

With a full set of equivalent widths and representative FWHM values, 
Equation 4 could be used to find a realistic uncertainty in a given line.  
These steps were repeated for all three CaT lines, then added in 
quadrature for the error in the summed equivalent width, $\Sigma$W.  
An additional term was added to this in quadrature, to account for
the variation in equivalent width measuring methods used.  This term
was determined from the rms dispersion about the best fit in the 
transformation to the \cite{Cole04} equivalent width measurement system (see $\S$3.1.1)
 
The final expression for the uncertainty in metallicity is now 
formulated into one equation, allowing for simple partial derivative 
based error propagation.  Defining an equation for metallicity based 
on the following equation of observables or calibrated variables:
\begin{center}
\begin{equation}
f = [Fe/H]_{CG97}=c_{1}\left[\Sigma W+\beta(V_{HB}-V)\right] - c_{2}
\end{equation}
\end{center}
allowed us to express the total uncertainty as:
\begin{center}
\begin{eqnarray}
\Delta[Fe/H]^{2} & = \left\lbrace\LARGE\left(\frac{\partial f}{\partial c_{1}}\LARGE\right)^{2}(\Delta c_{1})^{2} +
 \LARGE\left(\frac{\partial f}{\partial \Sigma W}\LARGE\right)^{2}(\Delta \Sigma W)^{2} \nonumber \right. \\
&\left. + \Large\left(\frac{\partial f}{\partial \beta}\Large\right)^{2}(\Delta \beta)^{2} + \ldots\right\rbrace\LARGE
\end{eqnarray}
\end{center}
The simple mathematical propagation is reasonable and justified, as the the variables involved in Equation (6) are uncorrelated and independent.  This propagation and error accounting results in an average uncertainty in metallicity of $\Delta$[Fe/H] = $\pm0.25$ dex.
\subsubsection{Three Line Justification}
We performed a parallel reduction omitting the weakest line in the calcium series (8498 ${\rm \AA}$).  The purpose of this exercise was to test whether the uncertainties in constructing the calcium index, $W'$, were significantly reduced with the rejection of the lowest signal source.  Composite two-line indices were created in order to compare the relative errors that occurred for exclusion of each line in the creation of the calcium index.  The \emph{average} relative error for the full 80 star sample on any of the line pairs, $\frac{\Delta W_{nm}}{W_{nm}}$ where $n$ and $m$ are the first, second or third lines of the Ca II triplet, were calculated to be $\frac{\Delta W_{23}}{W_{23}}=0.10$, $\frac{\Delta W_{13}}{W_{13}}=0.12$, and $\frac{\Delta W_{12}}{W_{12}}=0.11$ for the pairs.  This minimal deviation, suggests that inclusion of all three lines is warranted, because the equivalent width measures do not dominate the random error budget, and thus there is no benefit to dropping the 8498 ${\rm \AA}$ line.  

\subsection{Radial Velocity Measurements} 
Radial velocities were measured from the strong calcium lines which 
had previously had a wavelength dispersion solution applied from the sky 
OH lines.  The low signal to noise of the individual frames necessitated that
we perform the cross correlation radial velocity calculations on the combined
spectra, rather than each individual image.  As such, heliocentric velocity 
corrections were tailored to the individual exposures and applied 
prior to combining the spectra, due to the long temporal 
baseline (roughly four months) of the observations.  Once shifted and combined, the spectra were ready 
for radial velocity computation with the aid of template stars and a Fourier 
cross correlation routine (\textsc{fxcor}).  A total of 23 template radial 
velocity stars, observed with the same instrument setup, were used with the 
cross correlation routine. This computation provided error analysis 
automatically, with the median error in the heliocentric velocities being 
$\langle\delta V_{hel}\rangle = \pm 6$ km s$^{-1}$.  Systematic velocity 
errors  due to a star's position in the slit were removed by centroiding the stars relative to the slit centre.  The fact that this procedure was done on combined spectra resulted in small absolute corrections, as the $\surd n$ statistics meant that the individual slit errors were minimized in the combination and correction steps.  The typical shift for the slit error on an individual exposure is approximately 6-9 km s$^{-1}$, and we note that this shift produces negligible uncertainties in the equivalent width error.  The final average absolute corrections to the slit centering errors on the combined spectra were on the order of $\leq 1.5$ km s$^{-1}$.

Our sample of RGB stars has a mean velocity of 
$\langle V_{hel}\rangle = -130 \pm 1$ km s$^{-1}$, 
identical to the heliocentric velocity for WLM derived 
from neutral HI studies \cite{Jackson04, Kepley07}.
Only one foreground star was found, with radial velocity 
$V_{hel}$ = +48 km s$^{-1}$, leaving us with 78 stars total.  Given the location of WLM with respect to the Galactic plane and the colour and magnitude cuts in our preselection routine, it is
not surprising that there were so few foreground objects.

\subsection{Age Derivations}
An estimate of the relative ages of our sample of RGB stars was determined 
using the Victoria-Regina stellar evolution tracks \citep{VR06}. 
Given the extended SFH of WLM \citep{Mateo98, Dolphin00} and the gas rich 
nature of dIrr galaxies, we expect a significant range in red giant ages 
in the dataset.  

A CMD of the target stars in two metallicity bins, along 
with the fiducial sequences for M68 and 47Tuc, is shown in 
Figure \ref{fig:cmdtarg}.  As the stars were selected across a 
wide range in (V-I) colours, the spectroscopic CaT [Fe/H] estimates allow 
us to break the age-metallicity degeneracy and provide a more complete 
understanding of the evolved populations in WLM.   The age derivation 
procedure and error estimates of the relative stellar ages are more
fully described by \cite{Leaman08}\footnote{This thesis work is freely available for download at: http://hdl.handle.net/1828/1325}, 
however a brief outline is given below.

To determine an age from the Victoria-Regina models,  
a value of [Fe/H]$_{CaT}$, V, and (V-I) is required for each star, 
as well as global values for distance and reddening to WLM (see Table 1) 
and an assumption about the $\alpha$-element enhancement level. The [$\alpha$/Fe] value to use for the WLM sample is not clear especially given the age and metallicity range of the evolved populations here (e.g., dSph stars show [$\alpha$/Fe]$ = +0.3$ for metal-poor stars, but show subsolar ratios for increasing metallicity, \citealt{Tolstoy03, Venn08}). [$\alpha$/Fe]$ = +0.3$ was adopted based on the results from galactic globular clusters which were used to calibrate the CaT method here; however the effect of computing the ages with [$\alpha$/Fe]$ = 0.0$ was found to be minor.  Systematic effects (e.g., an evolutionary model's treatment of mixing 
length, convective overshooting, CNO abundances, radiative diffusion, 
gravitational settling etc.) between the models and observations were 
removed by comparing the isochrones to homogeneous globular cluster 
photometry from the CADC website\footnote{url: http://www3.cadc-ccda.hia-iha.nrc-cnrc.gc.ca/community/STETSON/standards/}.  
The location of each star in the dereddened plane was used to 
interpolate a best age estimate amongst tracks of the appropriate 
metallicity for that star.

RGB stars can only evolve to luminosities approaching the tip of the RGB when the core becomes significantly electron-degenerate before the onset of core helium burning.  Thus there is a lower limit to the age of the first-ascent RGB stars that could possibly be included in our sample, typically around 1.6 Gyr.  At a given metallicity, this age marks the blue edge of the colour distribution of RGB stars.  In cases where the star was bluer than allowed by this condition, an arbitrary age marker of 1 Gyr was assigned as an aid to interpreting the sample properties.  Similarly, those redward of the oldest available track (18 Gyr) were assigned an age of 
18 Gyr and a significant error (discussed below). 
As seen in Table 3, roughly one third of the stars are 
below the 1.6 Gyr cutoff, and most of those are found 
in the bar field of WLM.  

Formal errors on the ages were adopted by folding in the uncertainty 
in the spectroscopic metallicities, as well as the uncertainties in the 
photometry. 
While other astrophysical factors could impact the positions of the 
stars on the CMD (e.g., variations in relative internal reddening),  
we are unable to explicitly quantify these effects.  
Figure 13 of \cite{Cole05} illustrates the effect that some of these 
factors can have on the ages.  Given those dependencies and the value
of a typical metallicity error of $\pm 0.25$ dex, the random error in age is $\sim \pm 50 \%$.

The technique used here for deriving ages can be applied successfully in 
globular cluster populations in the MW \citep{Tolstoy03b}.  However studies using 
identical methods in dwarf galaxies have found it much more difficult 
to assign ages to evolved stars. \cite{Tolstoy03} first showed that the evolved stars in dSph galaxies often lay blueward or redward of the full age parameter space (see the discussion in $\S$3.2 \citealt{Tolstoy03}; also \citealt{Tolstoy03b}).  This phenomenon has also been seen in the LMC \citep{Cole05}.
This wide range of colours is clear in our Figure \ref{fig:cmdtarg}, where we overlay GGC fiducial sequences
(M68; [Fe/H]$\sim$-2.0, and 47 Tuc [Fe/H]$\sim$-0.7) 
onto a metallicity binned colour magnitude diagram of WLM.
Notable are both metal poor and metal rich WLM
stars\footnote{In this, and other representations throughout 
the paper, the metallicity split to characterize ``metal rich'' and 
``metal poor'' RGB stars is made at [Fe/H] = $-1.27$ dex - 
approximately the median and mean value of our sample.} 
blueward of the fiducial sequence for M68.  The majority of these stars are more metal rich than M68, suggesting that we are seeing a range of ages in our sample younger than typical globular cluster ages.  In Figure \ref{fig:cmdtarg} we also point out one star (STARID 19203), at (V-I)$_{0}\sim 2.5$, and note that a greater uncertainty should be applied to this star's parameters as it may be subject to excessive reddening, or spectral contamination (notably, from weak TiO bands).  

The question that we are left with is whether the bluest stars in our age derivations
(those with inferred ages $\leq$1.6 Gyr), are really intermediate-mass stars on the early-asymptotic giant branch or are true low-mass stars on the first-ascent RGB, influenced by uncertainties in composition or differential reddening.
If the majority of the bluest stars \emph{are} truly 1 Gyr or less, 
then it is reasonable to look for concentrations of their more-evolved descendants (e.g., thermally-pulsing AGB stars). 
Among Local Group dwarfs, WLM has been found to have an 
extraordinarily large ratio of carbon- to M-stars, C/M$=12.4\pm3.7$ \citep{Battinelli04}.  This large population of carbon stars, along with the high recent star formation of WLM, does suggest that the majority of stars in our youngest age bin are truly young.
Note that these young stars are not likely to be differentially reddened AGB or RSG contaminants, as cross-correlation with UKIRT WFCAM \textit{JHK} photometry rules this out.  However, early-AGB stars ascending from the horizontal branch towards the thermally-pulsing (carbon star) phase can occupy the same CMD space as a bona fide RGB star, and this becomes quite likely in a composite population such as WLM's.  We note that if an RGB star is in fact an early AGB star, at a given metallicity the age of that star would be $\sim 30\%$ older than we infer here (for further discussion see \citealt{Cole05}).

\section{Analysis and Discussion of WLM}
In this paper, we have determined the [Fe/H] and radial velocity 
values for 78 \emph{individual} RGB stars in WLM.    These can be used
to examine the structure, kinematics, and chemical evolution history of 
this galaxy for the first time in conjunction with derived age estimates.  
This is unique, as all previous 
studies of supergiant stars \citep{Venn04, Bresolin06, Urbaneja08} 
and HII regions \citep{Skillman89, Hodge95, Lee05}  sampled the young 
population and offer little insight into the earlier epochs of formation 
and evolution
of this galaxy.  Similarly, photometric studies \citep{Minniti97, Hodge99, 
Alan05} were only able to provide global views of the metal poor population, 
and were subject to degeneracies in age and metallicity.  

To begin our analysis, elliptical radii were determined for all of the stars 
in our sample.  The position angle of WLM was taken to be 181$^{\circ}$ 
\citep{Jackson04} and the eccentricity 0.59 \citep{Ables77}.  
After transforming the equatorial coordinates into the transverse 
$\xi$, $\eta$ plane, and calculating the axial components of the ellipse 
for a given star, the elliptical radius ($r_{ell}$) was determined from 
the geometric mean of the semi-major and semi-minor axes.  These elliptical radii have uncertainties due to the inclination with which WLM appears to observers (69$^{\circ}$, which is not assumed in the calculations; \citealt{Ables77}).  A given star may of course sit off the major plane of the galaxy - thus the calculated elliptical radii has an intrinsic degeneracy with the star's true distance from the plane of the disk.

A summary of the spatial and photometric properties for the 78 RGB stars
in this sample are listed in Table 3.   This includes the metallicity, 
age, radial velocity, and elliptical radii values as well.

\subsection{Chemistry}
The metallicity distribution function from our full sample of stars provides 
the first detailed insight into the chemistry of the evolved population of 
WLM.  The top panel of Figure \ref{fig:2pmdf} shows the MDF for the WLM sample 
subdivided into the bar and north fields.  The global mean ($-1.28 \pm 0.03$, 
$1\sigma = 0.37$)\footnote{It should be noted our observed spread in metallicities should be compared to the intrinsic uncertainty in obtaining our individual [Fe/H] estimates (i.e., $\Delta$[Fe/H]=$\pm 0.25$ dex)} and peak ($\sim -1.30$) values of the MDF are slightly 
higher than the photometric determinations, but consistent within errors.
For example, \cite{Hodge99} derived [Fe/H] = $-1.51 \pm 0.09$ by isochrone 
fitting to the lone GC in WLM, \cite{Alan05} found [Fe/H] =$ -1.5 \pm 0.2$ 
(assuming no alpha-element 
enhancements) from empirically calibrated RGB colours, and  
\cite{Minniti97} found [Fe/H] = $-1.45 \pm 0.2$, again from the photometric 
properties of the RGB. These photometric metallicity estimates essentially give lower limits,
as younger stars will be shifted bluer on the RGB.  Our higher results confirm that effect - as shown
by our higher metallicities and range of ages for this composite population.

The MDF in Figure \ref{fig:2pmdf} shows a sharp drop off at the low end, with very few metal poor stars below [Fe/H] $\leq -1.8$. 
\cite{Dolphin00} suggested that there was a significant star formation event
9-12 Gyr ago with an enrichment up to [Fe/H]$_{0} = -2.18 \pm 0.28$, 
however those stars do not appear in our sample.  The low 
metallicity dropoff is primarily due to observational difficulties in attempting to sample a sizeable relative number of low [Fe/H] stars.  Metal poor stars at this evolutionary locus are difficult to find in any galaxy due to a population bias resulting from mass dependent stellar evolutionary timescales (c.f. \citealt{Cole08}).  For a given time interval, more young, high mass, high metallicity stars will evolve to the TRGB than older, low mass, metal poor stars in the same time interval.  The dominant fraction of young stars at the TRGB is large enough to outnumber the metal poor old stars, despite the contradictory initial relative numbers of the IMF.  This effect will be enhanced in the presence of any age-metallicity relationship and is still present for populations with extended SFHs like WLM.  In order to quantify the probability 
that the metallicities in our sample are drawn from a Gaussian parent 
distribution (which we don't expect due to the extended SFH, but wish to quantify), we performed a Kolmogorov-Smirnov (K-S) test on the cumulative 
metallicity distribution function (CMDF).  The probability that the distribution was Gaussian is $\sim 8 \times 10^{-19}$.  More interesting is a comparison to a metallicity distribution predicted by a simple closed or leaky box chemical evolution model.  The bottom panel of Figure \ref{fig:2pmdf} shows our cumulative metallicity function compared to predicted distributions from leaky box models with a range of values for the parameter `c', which controls the effective yield.  The K-S probabilities that our distribution is drawn from one of the simple one zone models is $\leq 4\%$ for all models shown.  Our distribution clearly shows a lower number of metal poor stars compared to these models.  

Not surprisingly, the young stars that make up the metal rich half of our sample are found preferentially in the bar field of WLM (see Figure \ref{fig:2pmdf} top panel).  
Several tracers of high active SF are observed in this inner field, notably 
the strong and numerous HII regions \citep{Hodge95, Youngblood99} 
consistent again with the most enriched material being found at low r$_{ell}$.

This is more clearly visualized in the top panel of Figure \ref{fig:2prell}, where the mean metallicity decreases by 
$\sim$0.3 dex with distance from the galactic core.  From that figure we find a radial metallicity gradient of 
$\frac{\partial[Fe/H]}{\partial r}= -0.14 \pm 0.02$ dex/kpc.  This smooth gradient may however be a line of sight superposition of two (or more) spatially localized subpopulations, mimicking a smooth radial profile. 

This population change is consistent with the results from the photometric study 
by \cite{Minniti97}.  The [Fe/H] values show a dispersion in metallicity 
that is roughly constant over the sampled region of the galaxy 
($\sigma$ = 0.32), similar to previous analyses.  A two sided Kolmogorov-Smirnov test of the bar and north field cumulative metallicity distributions reveals that they have a very small probability of being drawn from the same parent distribution ($P=0.005$).  Splitting our sample into those stars with $r_{ell} \leq 0.1$ deg. and $r_{ell} > 0.1$ deg. and running the same K-S test, results in a similar probability ($P=0.024$).  This would appear to strengthen the interpretation of distinct spatial populations in WLM, as the inner and outer CMDFs show statistically different properties - not just different mean metallicities.  

The bottom panel of Figure \ref{fig:2prell} shows the cumulative radial 
number density (CDF) of metal poor and metal rich stars as a function of 
elliptical radius from the center of WLM.  The metal poor stars are 
found in more uniform numbers independent of radial position within 
the galaxy, whereas the more metal enhanced stars are found 
preferentially closer to the center of WLM.

In Figure \ref{fig:ymolb}, we compare the unbinned CMDFs for three age bins of our sample.  In each bin, we have compared the WLM subsample to predicted CMDFs from a leaky, and closed box chemical evolution model.  While our sample deviates from the simple models in the youngest age bins, the oldest age bin ($t > 6$ Gyr) shows a large probability of being drawn from a distribution approximated by the closed box model.  We might naively expect that our oldest stars would have a characteristic shape to their metallicity distribution that would be more closely approximated by a closed box scenario, where little infall or outflow occurs.  The deviations at younger ages presumably occur with the increase of accreted gas, and as such, the CMDF of our younger stars begins to vary from the shape predicted by a closed box model.  However, there is an additional astrophysical effect that must be considered.  As mentioned in paragraph two of this section, the evolutionary flux through the TRGB results in higher numbers of young stars in that locus than older, low mass, metal poor stars.  This effect means that our 78 star sample is a slightly skewed example of the true [Fe/H] distribution at the TRGB.  As such, one should keep in mind that the probabilities found in the K-S tests are really describing how our sample is comparing to the closed box models - presumably the true metallicity distribution of WLM would have more metal poor stars, and the K-S probabilities would be higher.  This population bias is much more prominent in the youngest age bin, and at the older bin is less of a factor, and so these stars may result in a more representative sample of WLM's metallicity distribution.  With larger numbers of stars it will be possible to quantify how much of the probability change with age is due to an evolutionary population bias, and how much may be attributed to infall of pristine gas.

\subsection{Chemodynamics}
Figure \ref{fig:2fvdf} shows the results of our radial 
velocity analysis, presented as two velocity distribution functions (VDFs) 
for the bar and north field.  Each field has been further subdivided into 
metal poor and metal rich subsamples. We note the peaked, Gaussian 
distributions around the mean values of 
$\langle v_{north} \rangle = -136 \pm 3$ km s$^{-1}$ 
and $\langle v_{bar} \rangle = -126 \pm 3$ km s$^{-1}$ for the north and bar 
field samples respectively.  Comparing these subsamples shows that the 
{\it mean velocities} 
of the north and bar, metal poor and metal rich, populations are 
consistent (i.e., $\langle v_{NMP} \rangle = 
\langle v_{NMR} \rangle = -136 \pm 3$ km s$^{-1}$; 
$\langle v_{BMP} \rangle = 
\langle v_{BMR} \rangle = -126 \pm 4$ km s$^{-1}$).  
This indicates that the general bulk motion of both subgroups of stars 
in a given field holds in accordance with the group mean.  
However the {\it velocity dispersions} of the metallicity subpopulations,
do not show the same uniform characteristics.  The velocity dispersions 
of the metal poor subsamples for the north and bar fields are 
larger than the metal rich samples in both fields.  We find: 
$\sigma_{v(NMP)} = 19$ km s$^{-1}$, 
$\sigma_{v(NMR)} = 12$ km s$^{-1}$, $\sigma_{v(BMP)} = 20$ km s$^{-1}$, 
and $\sigma_{v(BMR)} = 14$ km s$^{-1}$.  
The Sculptor dSph galaxy \citep{Tolstoy04} shows similar chemodynamic trends.

Figure \ref{fig:vvmet} again illustrates that the 
velocity and metallicity of the stars in the WLM sample are changing 
with respect to position in the galaxy.  The north field stars are
more metal poor and have a higher approach velocity relative to the bar 
field stars.  This chemodynamic signature is consistent with the 
interpretation of previous photometric studies that found a
population gradient in WLM.
However, our results separate the metallicity-velocity-position
degeneracies, and we have also determined ages.    

\subsection{Properties of Each Stellar Population and 
 Comparisons to the Gas Dynamics}

The sample is divided into three age bins to examine
the means and dispersions in the metallicities, velocities,
ages, and elliptical radii of each stellar population, as listed in Table 4. 
Generally, the older stars are more metal poor and kinematically hotter 
than the younger stars.  It should be stressed however, that 
the primary variations we see in our data are \emph{radial} velocity and 
metallicity changes.  This is confirmed unambiguously by a 4-d 
principal component analysis, and a k-means clustering analysis 
(see \citealt{Leaman08} $\S$4.5).

Of equal interest is the examination of the properties of the stars 
which do not show the same kinematic traits as the gas (from HI studies).  
The simplest expectation is that the metal rich stars would have
formed from material that is more recently dynamically coupled to 
the gas, and would have had less time to become kinematically hot.  
This is borne out by the smaller velocity dispersion among the 
metal rich sub-sample of our dataset (see $\S$4.2).

In addition to this, we can compare the stellar velocity
gradient to that of the HI velocity gradient.
\cite{Jackson04} found an HI velocity gradient consistent 
with a rotation curve, and \cite{Kepley07} derived a rotation velocity 
of $\sim$ 30 km s$^{-1}$ for the gas, with a more steeply varying 
velocity gradient in the northern approaching side of the galaxy.  
A direct comparison of the stellar and gas kinematics is shown in Figure \ref{fig:pvnp}, where the \cite{Kepley07} HI position-velocity curve is overlaid on our stellar velocities.  The average velocity 
separation between the flat central part of our stellar data 
and the approaching northern section is $\sim 15$ km s$^{-1}$.  This 
is statistically significant given our errors of $\pm 6$ km s$^{-1}$ 
per star and error on the mean of $\pm$ 3 km s$^{-1}$.  However the 
gradient is shallower and only half that of the gas rotation velocity.  
An apparent spatial lag is also seen, although the rotation center 
is consistent with the HI studies.  These types of stellar and gas velocity discrepancies may be common in dwarf irregular galaxies, whether isolated or interacting, as witnessed by the massive star kinematics in the Small Magellanic Cloud reported by \cite{Evans08}

Clearly the stellar members in this sample are not in the same velocity space as the bulk of the rotating gaseous component of WLM.  In the northern part of the galaxy, the stars lag the gas rotation by as much as $\sim 35$ km s$^{-1}$.  At the southern most part of our fields, the stars show a velocity offset of roughly the same magnitude, but opposite sign.  To fully explain the complex velocity decoupling seen between the stellar and gaseous components of WLM will take a larger sample, but Figure \ref{fig:pvnp} nicely illustrates the lack of coherent motion.

The parameter $\Delta V_{(star-gas)}$ has been calculated to quantify the
difference in velocity of the stars from the gaseous components.  
For each star an offset in velocity was found 
by subtracting the stars' velocity from that of the gas velocity at the same 
right ascension and declination in WLM, according to an HI velocity map provided by Dr. A. Kepley (priv. comm.).  This was done to search for subsamples of stars
that track the gas velocity the closest.  
If the HI gas in WLM forms a thin rotating plane, then the youngest 
stars in our sample may track this better, than say, the oldest stars
In Figure \ref{fig:gdelt}, 
$\Delta V_{(star-gas)}$ is plotted against $v_{hel}$, age, and metallicity.  
While there is some substructure in these representations, no clear trends 
emerge.  This suggests that the stellar occupation of the HI velocity space 
is essentially random - that all stellar age bins/populations have been 
dynamically excited from the HI bulk motion.  
This would be consistent with the stellar population of WLM laying in a 
thick disk and/or spheroidal configuration.  Even the youngest stars show 
a range in $\Delta V_{(star-gas)}$, thus either the gas and stars were never 
coupled (e.g., if the gas was recently accreted) or the mechanism 
responsible for dynamically decoupling the stellar velocities from the gas 
must have been present within the last few Gyr.

A bar in WLM's recent history could provide a natural explanation to the 
lack of coherent chemistry and ages amongst the stars that track the gas velocities the closest (recall Figure \ref{fig:pvnp}).
Bars will dynamically stir a stellar population in a low mass galaxy such 
as WLM, and this also offers an explanation for the large velocity 
dispersions and velocity offsets in our sample when compared to the gas.  
\cite{Olsen07} noted tidal heating may be responsible for a dynamically hot 
component in the stellar disk of the LMC.  However, given the isolation of 
WLM, a past bar may be a more feasible explanation for the dynamically hot 
component of WLM's stellar body.  

Small scale HI features, including a bar, have been investigated by 
\cite{Kepley07} and \cite{Jackson04}.  Both identify a centrally 
evacuated region in the disk.  
\cite{Kepley07} identify this as a ring or ``hook'' feature related to
concentrated star formation and subsequent supernovae explosions, while
{\cite{Jackson04} suggest either a central bar \emph{or} a massive 
blowout region.  
\cite{Kepley07} place an upper limit on the age of the ring at 128 Myr,
which is much younger than the metal rich, young stars in our sample 
($\leq 1.6$ Gyr).  However, these young stars also show a statistically 
significant peak in their radial velocity dispersion profile 
at the extent of the ring feature (see \citealt{Leaman08}).  
If these metal rich (and youngest) stars are responding dynamically to this 
feature, then this suggests the underlying physical mechanism is a resonance with longer
term effects than sampled by the gas features alone.
Interestingly, \cite{Kepley07} estimate that as much as 20\% of the 
total mass of WLM may be associated with this gas feature.  While \cite{Kepley07} note the HI isovelocity contours do not show the characteristic `s' shape of a bar, nor is the WLM IRAC data suggestive of one - the resonances affecting older stellar populations set up up by a past bar may outlive either of those indicators.

The ratio of rotation velocity to velocity dispersion 
($\frac{v_{rot}}{\sigma_{v}}$) can provide information on the 
amount that a galaxy's structure is rotationally supported.  In Table 5, 
we compare this ratio for our full sample of RGB stars, and various 
subpopulations.  The stellar velocity ratio for the full sample is lower than 
that derived for the gas kinematics in \cite{Mateo98},  although the gas 
velocity dispersion and the central density were assumed quantities.  
Our stellar rotational velocities were estimated by linearly fitting the 
heliocentric velocities versus elliptical radii.  
As expected, the metal poor, old, and ``halo-like'' subdivisions produce 
lower relative $\frac{v_{rot}}{\sigma_{v}}$ ratios than the other splits. 
These results suggest that the older stellar population in WLM is more pressure 
supported, while the younger stellar population (and gas) is more rotationally
supported.  While \emph{all} of our stars lay off the plane occupied by the gas in position-velocity space, and appear dynamically excited and decoupled from the gas motion - we see a range of $\frac{v_{rot}}{\sigma_{v}}$ for the stellar subpopulations \emph{within} that thick disk of stars itself. 

While all of these results are intriguing, we stress that this work
is limited by a very small statistical sample, particularly our last result
since the older stars that suggest that stellar population is
pressure supported rests on only 13 individual RGB stars with t $>$ 6 Gyr.  
A larger sample of stars is necessary to examine this and other results,
and ultimately test the conditions of the isolated dwarf galaxies from 
the earliest epochs of star formation to the present.  We have shown with this pilot
survey that spectroscopic analysis of evolved WLM stars is possible, and with future observations
on 8m class telesecopes we will be able to increase the sample size siginifcantly
with additional fields.

\begin{tiny}\end{tiny}

\section{Summary}
We have presented the first detailed analysis of the Ca II triplet spectra 
of 78 individual RGB stars in the isolated dwarf irregular galaxy WLM.
Medium resolution spectra taken with the FORS2 spectrograph at the VLT 
in two fields in this galaxy (a central bar region, and a 
northern field) were coadded to improve the SNR $\ge$ 20.
This is the first analysis of such faint RGB stars for \emph{both} 
velocity and metallicity measurements, and therefore we have been 
especially careful in our data reduction techniques and error propagation 
calculations.
The equivalent widths were converted into metallicities for each star
using the calibration from \cite{Cole04} with final errors $\sim \pm 0.25$ dex per star. 
The heliocentric velocities were determined to within $\pm 6$ km s$^{-1}$ (random error).  Relative ages were derived from theoretical stellar evolution models, however differential extinction and evolutionary effects may bias our age distribution.
 
The north field and bar field metallicity distributions have mean 
metallicities and dispersions of [Fe/H]$= -1.45\pm0.04$, $\sigma=0.33$ from 34 stars, 
and [Fe/H]$= -1.14\pm0.04$, $\sigma=0.39$ from 
44 stars, respectively.  Both metallicity distributions are narrowly 
constrained with a range of approximately $\Delta$[Fe/H]$\simeq 1.0$.
The stars in our sample show radial metallicity and velocity gradients, 
but interestingly the stellar rotation velocity is only approximately 
half that of the gas.  The stellar velocities appear to be kinematically decoupled and in a
thicker disk than the gas.  Detailed comparisons with the HI gas, along with the isolation of WLM,
suggest that the gas and stars were never coupled (e.g., recent HI infall) or 
that the stellar dynamics have been affected by resonances associated with a past bar.

WLM shows signatures of a metal poor population that is spatially and temporally distinct, and kinematically hotter than the young population at all radii - perhaps indicative of an extended halo.  The velocities of the older stars suggest that this evolved stellar population in WLM is pressure supported, as opposed to the younger stars
and gas disk which appear more rotationally supported.   This latter result
rests on a very small number of old RGB stars (13 with t $\ge$ 6 Gyr),
but could have significant consequences for constraining theories of
the structure and dynamics of dwarf galaxies in the early Universe.
A larger sample and survey size is necessary to confirm these results
and further characterize the stellar populations in WLM.

\acknowledgments
The authors thank Dr. Don VandenBerg for many useful discussions, and are grateful to Dr. Carme Gallart for many helpful comments on the original manuscript. We would like to thank Dr. Amanda Kepley for generously sharing her HI data. RL acknowledges support from NSERC Discovery Grants to DV and KV. RL also thanks SS for keeping him sane and grounded. 


\clearpage

\begin{figure}
\begin{center}
\plotone{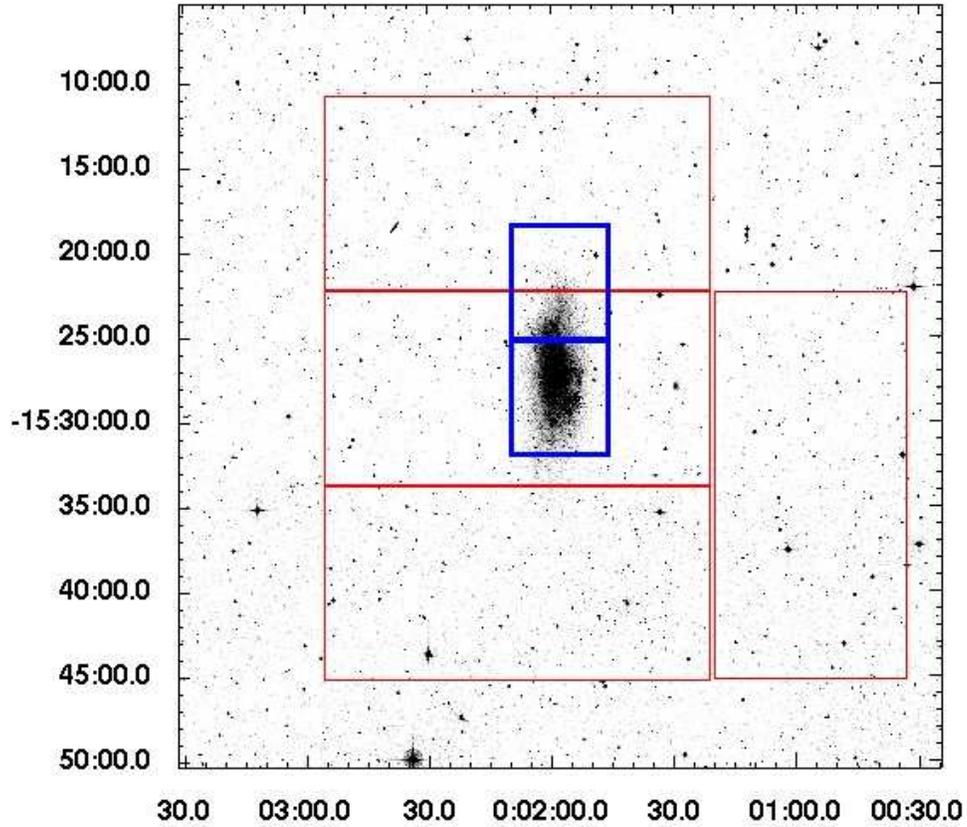}
\caption{Digitized Sky Survey SERC-J image of WLM.  The total image is 
approximately $45'\times45'$, with North being up and East to the left.
The relative locations of the north and bar fields (\textit{blue}) 
from this FORS2 spectroscopic work are shown.  The four red boxes 
indicate the fields imaged by the INT WFC survey (see \citealt{Alan05})  The two A-type supergiants 
from \cite{Venn03} are located approximately in the centre of our bar field, along with the HII regions from \cite{Hodge95}, and the B supergiants from \cite{Bresolin06}.  
}
\label{fig:fov}
\end{center}
\end{figure}

\clearpage

\begin{figure}
\begin{center}
\includegraphics[angle=90,width=1.0\textwidth]{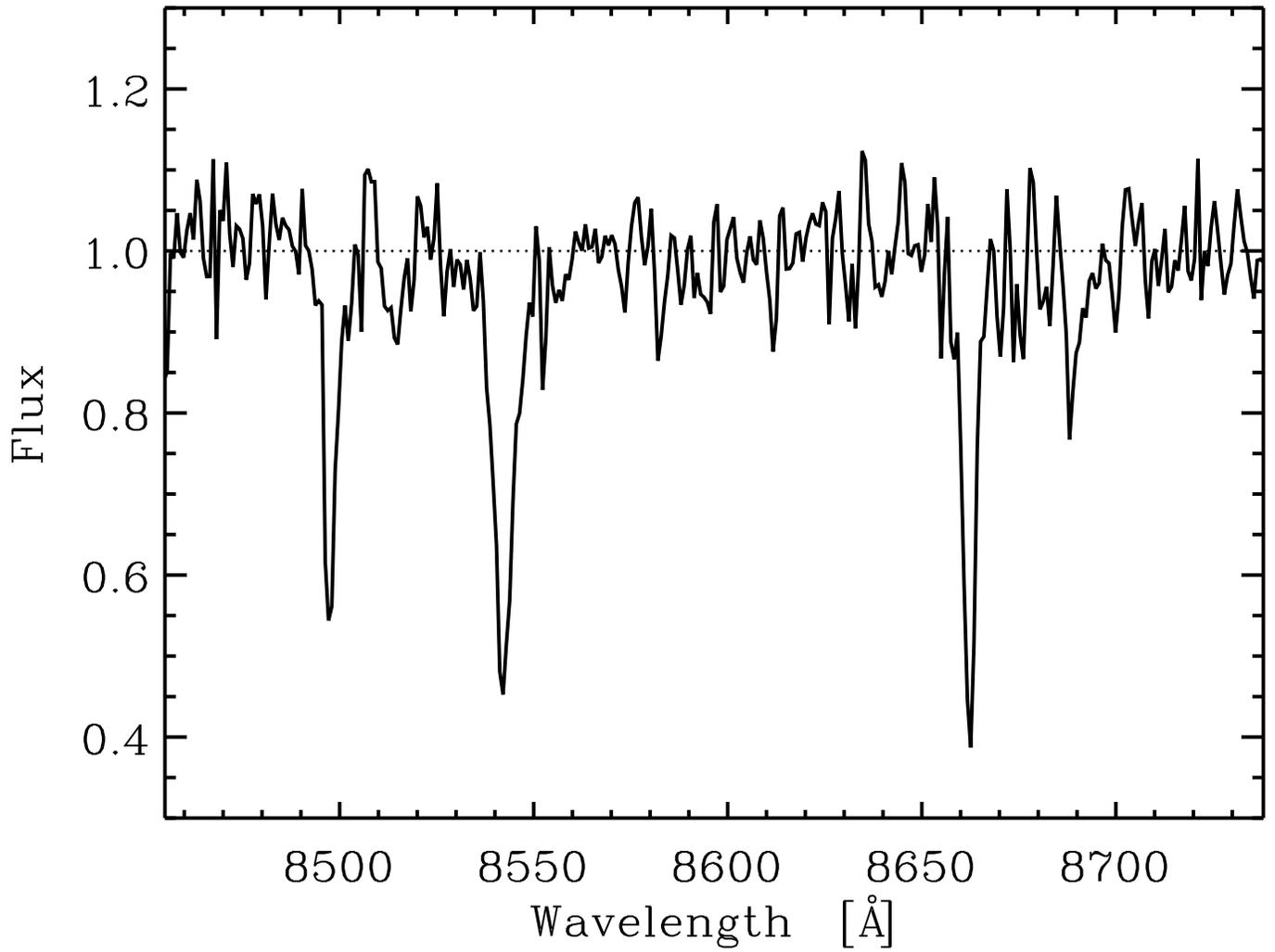}
\caption{Sample combined spectra for one star (STARID 28531) in the WLM FORS2 north field 
(\textit{top left}).  The prominent calcium absorption features are 
visible at $\lambda\lambda\sim$ 8498${\rm \AA}$, 8542${\rm \AA}$, 
and 8662 ${\rm \AA}$ even in this (S/N)$\sim$22 image.  The y-axis are relative flux units.  A dotted line has been drawn at the global continuum level of unity.}
\label{fig:fourspec}
\end{center}
\end{figure}

\clearpage

\begin{figure}
\begin{center}
\plotone{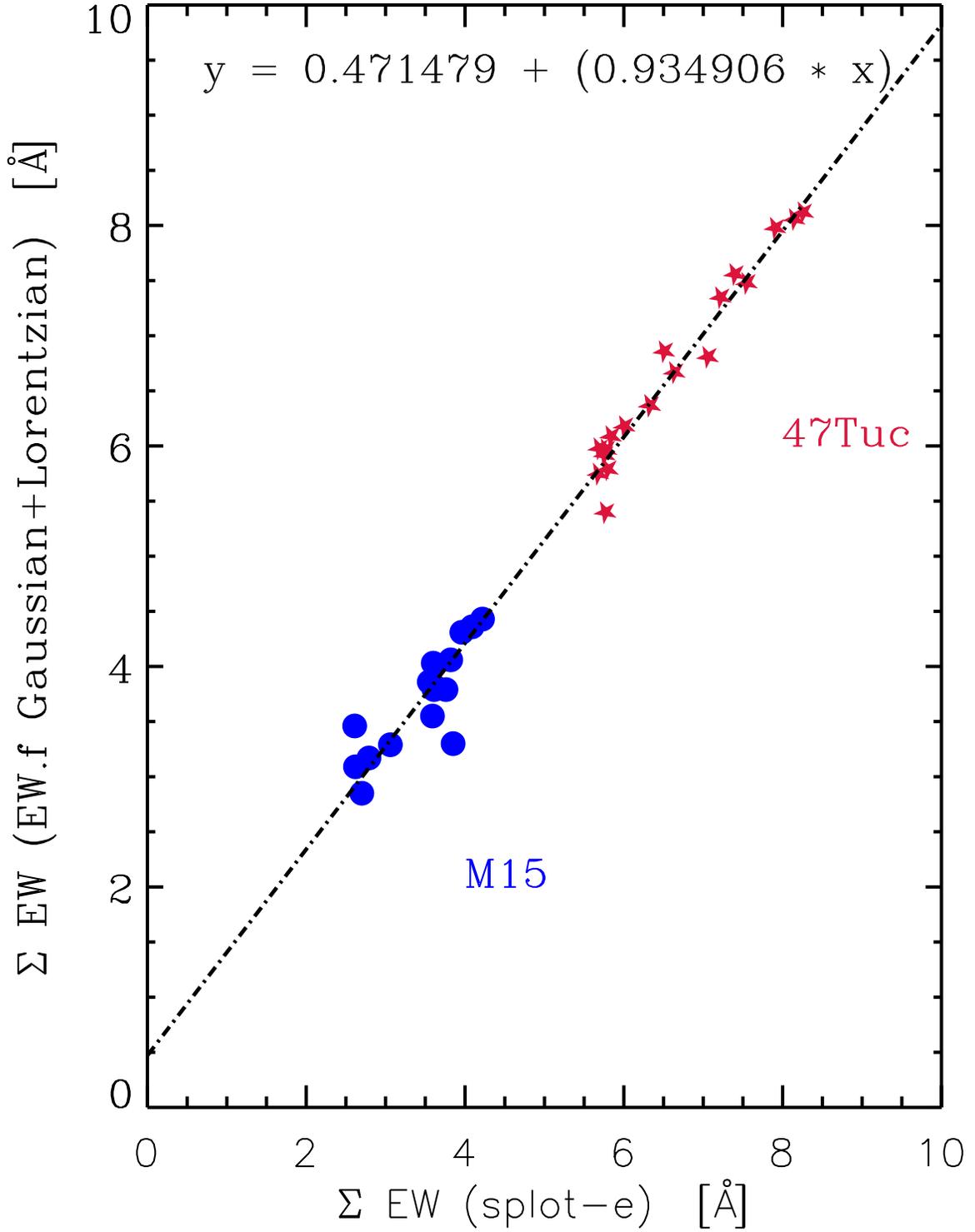}
\caption{Comparison of $\Sigma$EW measurements using the pixel integration method described in this work, and the Gaussian + Lorentzian profile fits from \cite{Cole04}, for two extreme metal poor and rich calibrating galactic globular clusters.  The results of a global linear least squares fit to the two cluster stars is shown as the dot dashed line.  The rms dispersion about the fit is 0.22 ${\rm \AA}$.}
\label{fig:sigewcomp}
\end{center}
\end{figure}

\clearpage

\begin{figure}
\begin{center}
\plotone{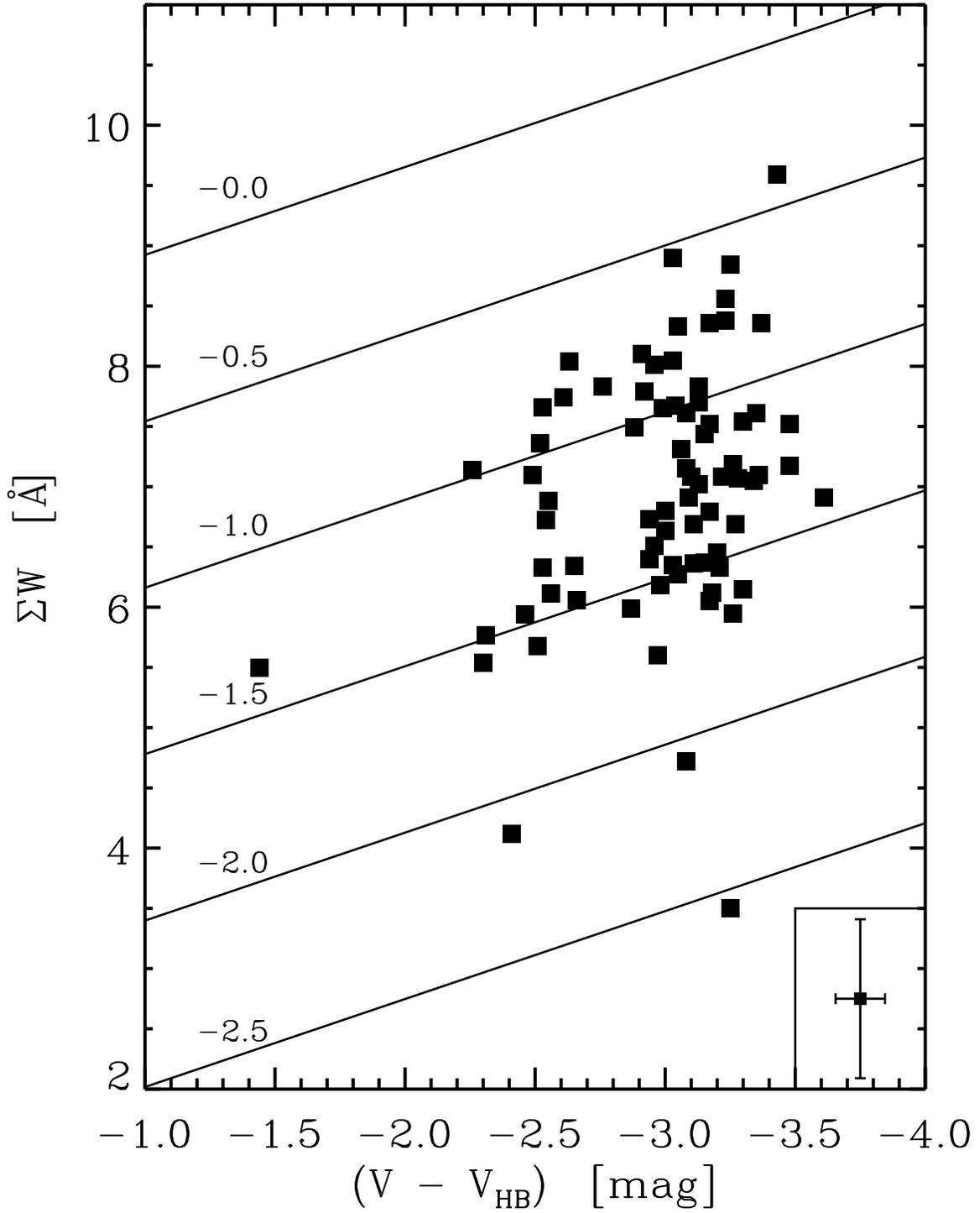}
\caption{Summed equivalent width of the Calcium II triplet lines 
($\Sigma$W) versus the V magnitude above the horizontal branch.  
The solid lines show constant metallicity according to the calibration 
by \cite{Cole04} which we used to derive our final [Fe/H] values (see text).}
\label{fig:vvsigewfl}
\end{center}
\end{figure}

\clearpage

\begin{figure}
\begin{center}
\plotone{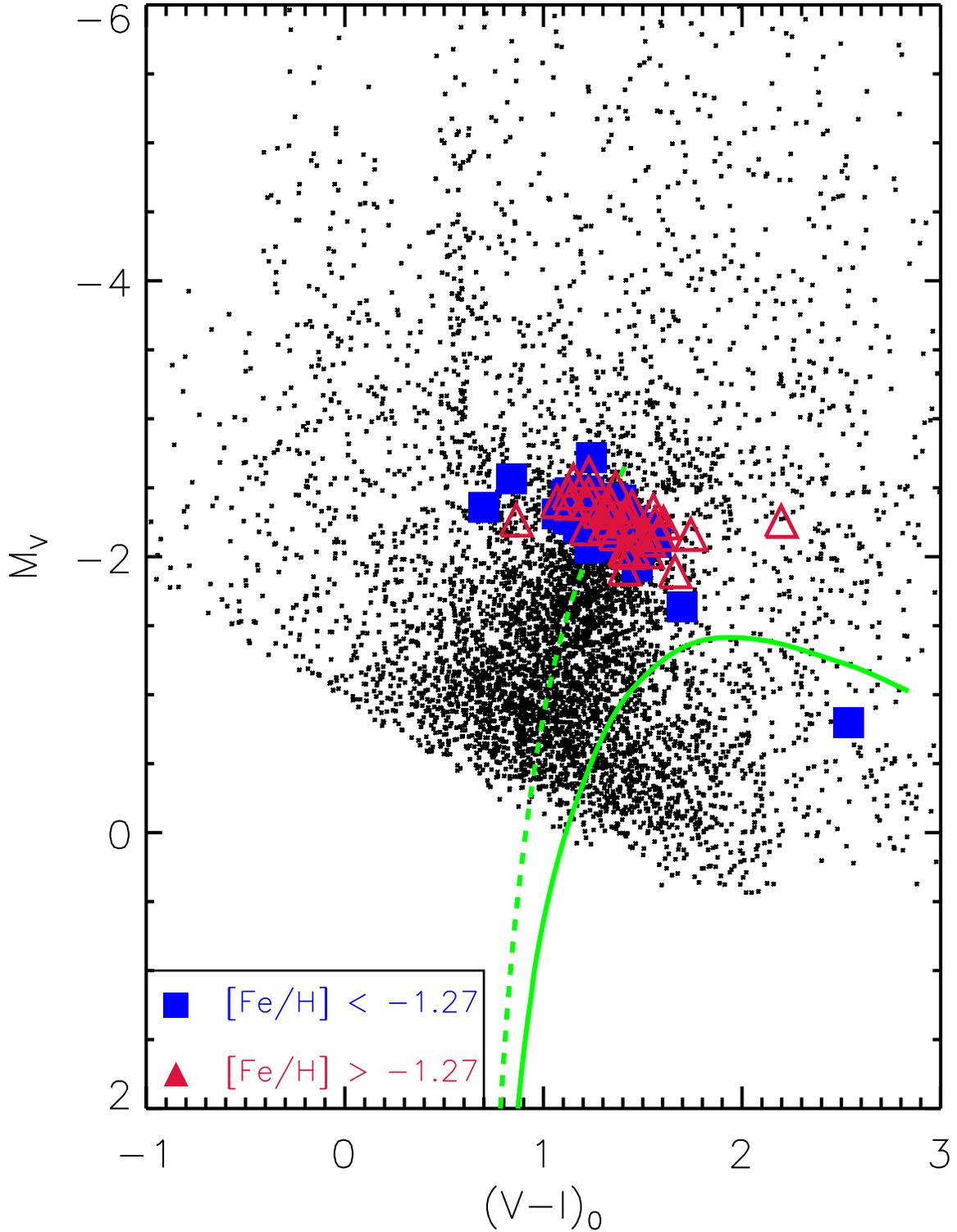}
\caption{CMD of target stars with [Fe/H] $\leq -1.27$ (\textit{blue boxes}) 
and $\geq$ -1.27 (\textit{red triangles}) with the fiducial sequences for 
47 Tuc ([Fe/H] $\sim$ -0.7) and M68 ([Fe/H]$\sim$ -2.0 \emph{green dashed line}).  The full catalogue of INT WFC photometry is shown as the black dots.  It can be
seen that there are both metal poor and metal rich stars blueward of the 
fiducial sequence for M68.  This indicates that we are most likely sampling 
a younger population in some areas.  More information on age trends can
be found in $\S$4.1.}
\label{fig:cmdtarg}
\end{center}
\end{figure}

\clearpage

\begin{figure}
\begin{center}
\plotone{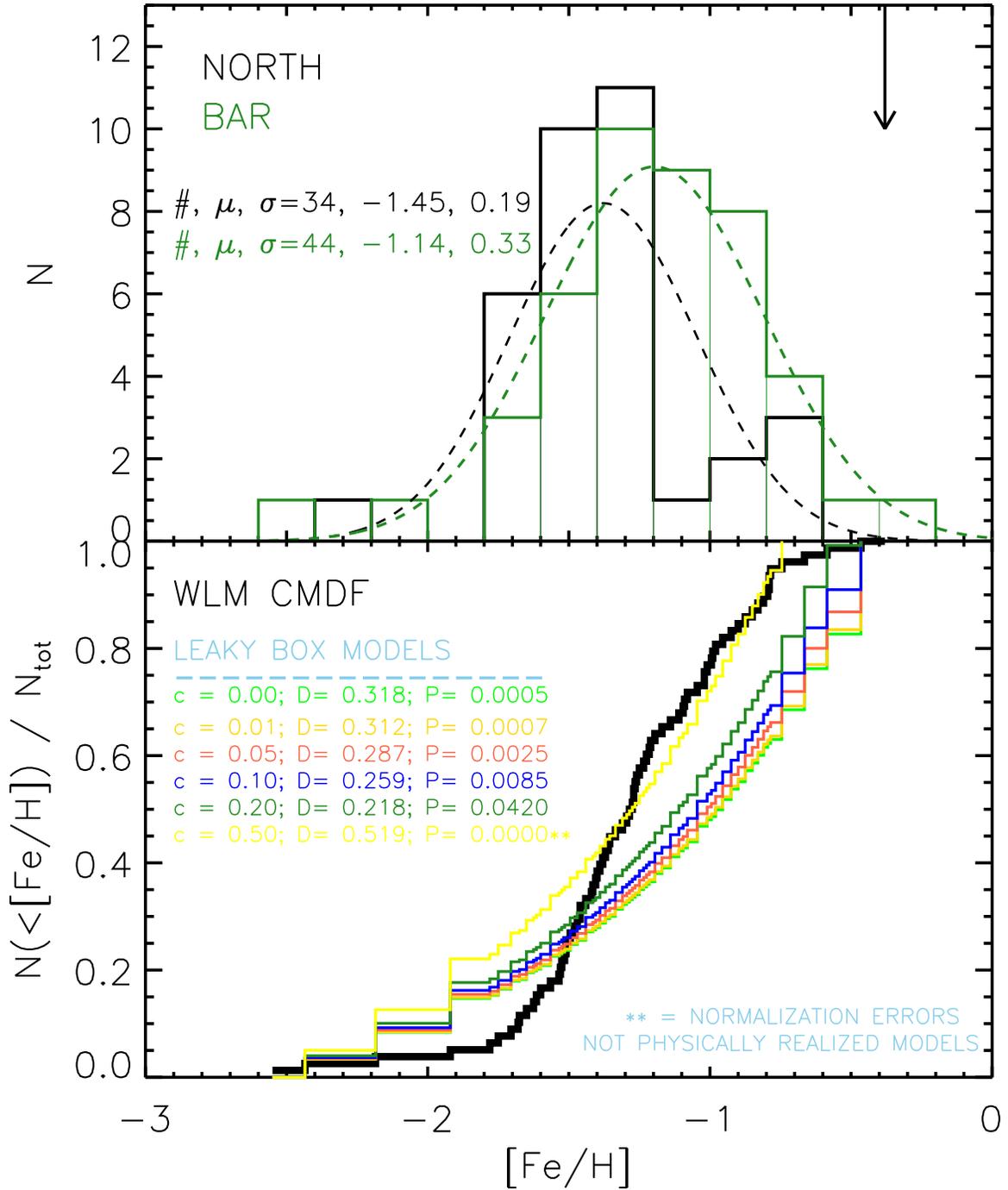}
\caption{
The metallicity distribution function for the two FORS2/MXU fields in WLM
(\textit{top panel}).  The bin size is 0.2 dex, and fitted Gaussians with 
parameters shown are indicated by the dashed lines.  Note that the mean 
values for the two fields differ by $\sim 0.3$ dex, and there is an absence 
of extreme or even moderately metal poor stars in either field.  The [Fe/H] estimates
from the \cite{Venn03} supergiants are shown by the solid arrow.
The cumulative metallicity distribution function 
(unbinned) for our WLM sample is shown in the bottom panel.  
The coloured lines represent cumulative distributions for simple leaky box chemical evolution
models, with different values (\emph{c}) controlling the effective yield).  Corresponding Kolmogorov-Smirnov D-statistics (\emph{D}), and probabilities (\emph{P}) are shown as well.}
\label{fig:2pmdf}
\end{center}
\end{figure}

\clearpage

\begin{figure}
\begin{center}
\plotone{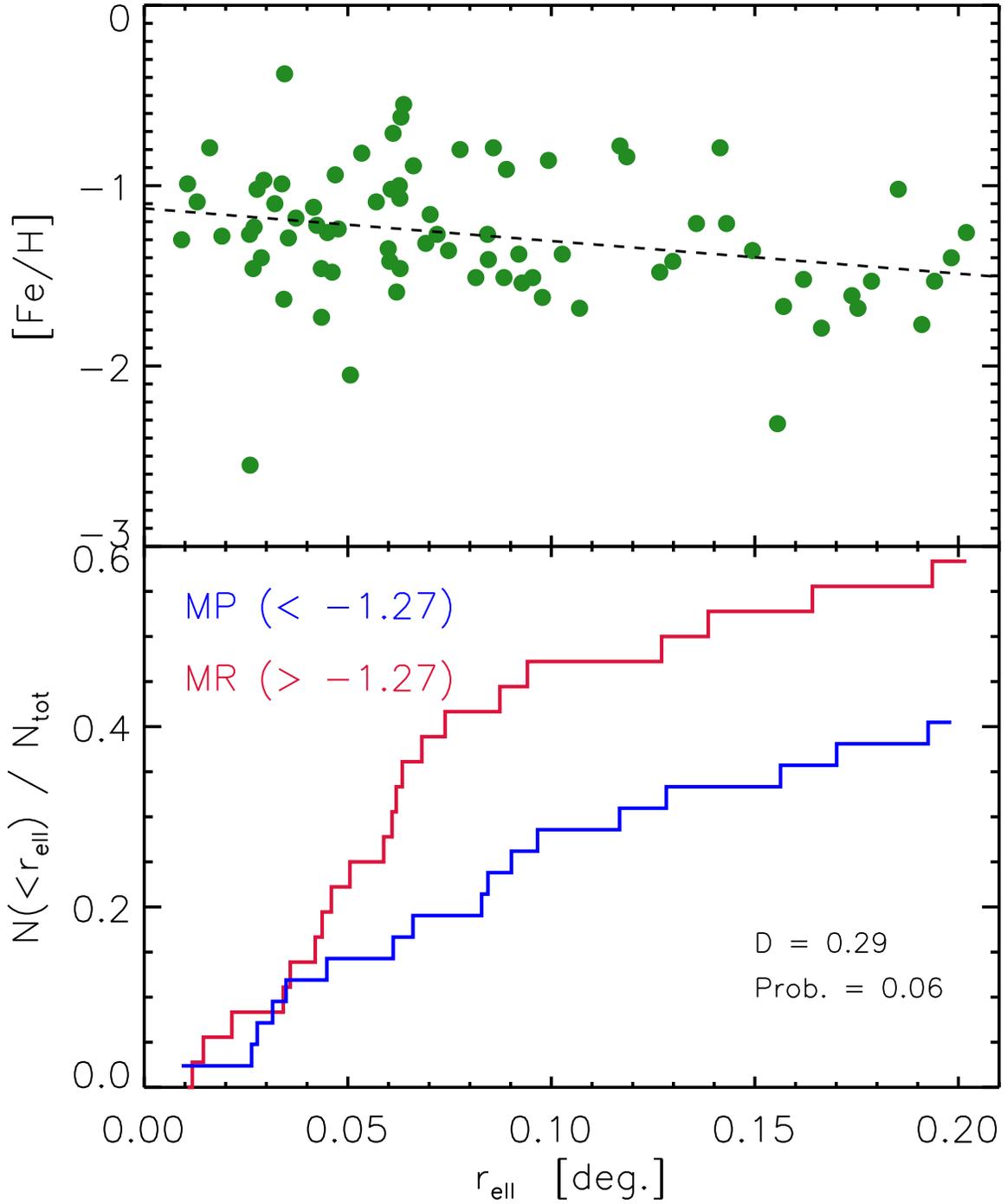}
\caption{[Fe/H] vs. elliptical radii for the full sample of RGB stars in WLM 
(\textit{top panel}).  A radial change is clearly evident, 
and quantitatively confirms the interpretation of the disk of WLM 
being a region of higher SF in more recent times relative to the outer 
galactic regions.  
Cumulative distribution functions vs. elliptical radii for the 
metal rich and metal poor subsamples (\textit{bottom panel}).  
The metal poor half of the sample rises more linearly over 
the radial extent of our survey, compared to the more centrally 
concentrated metal rich subpopulation. Shown are the K-S D statistic, and probability for a two sided K-S test of the two distributions.}
\label{fig:2prell}
\end{center}
\end{figure}

\clearpage

\begin{figure}
\begin{center}
\includegraphics[angle=90,width=1.0\textwidth]{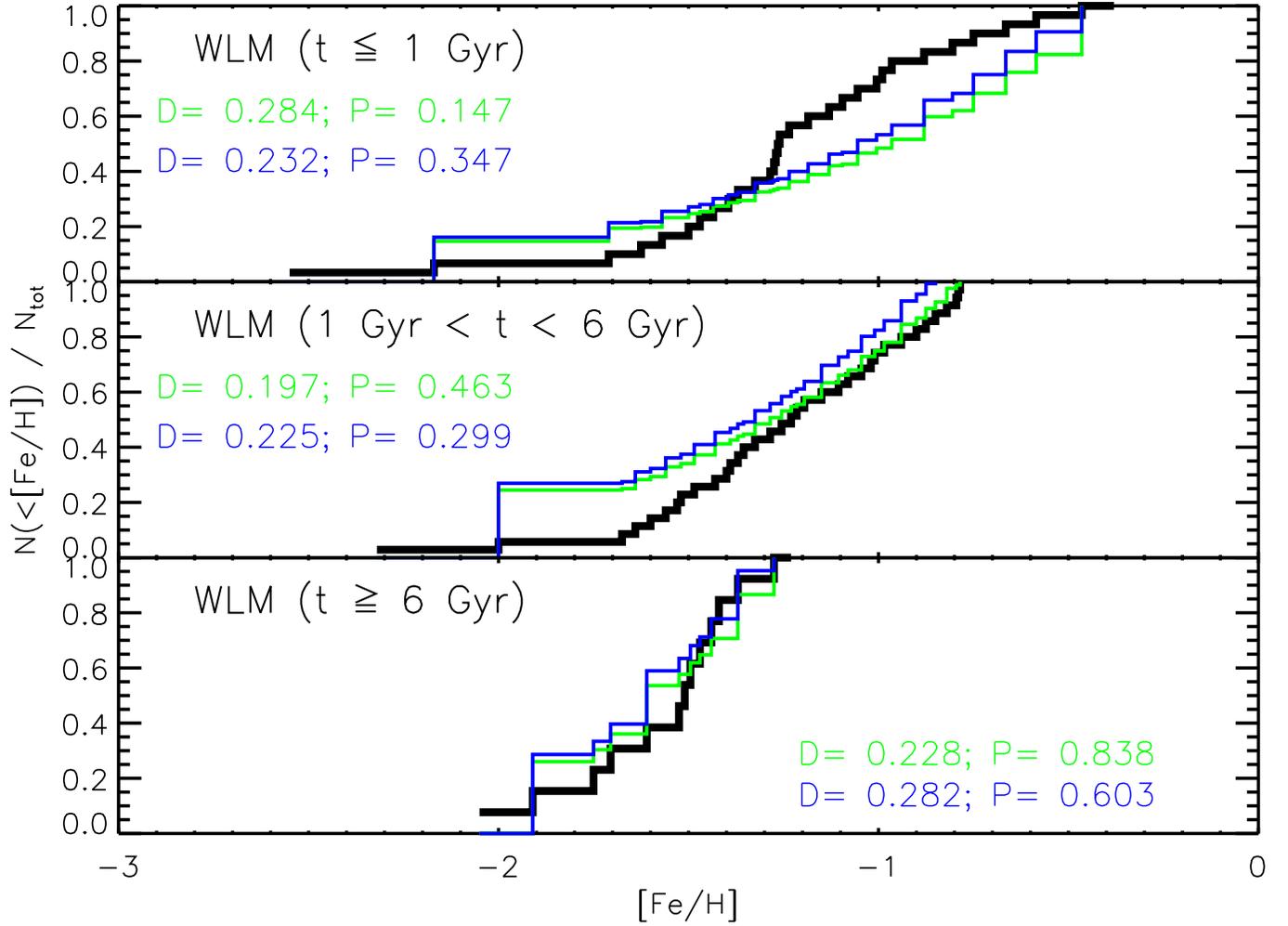}
\caption{Cumulative metallicity distributions for our WLM sample, subdivided by ages: $t < 1$ Gyr (\emph{top panel}), $1 \leq t \leq 6$ (\emph{middle panel}), $t > 6$ Gyr (\emph{bottom panel}).  Comparison CDFs based on simple closed (\emph{green}), and leaky box (\emph{blue}) chemical evolution models are shown for each age set.  The two sided Kolmogorov-Smirnov D statistics are shown, as well as the probability that our metallicity distributions are drawn from one of the theoretical distributions.  The probability rises steeply with age for the closed box model, with $P \sim 14\%, 46\%, 84\%$ in the three bins.}
\label{fig:ymolb}
\end{center}
\end{figure}

\clearpage

\begin{figure}
\begin{center}
\plotone{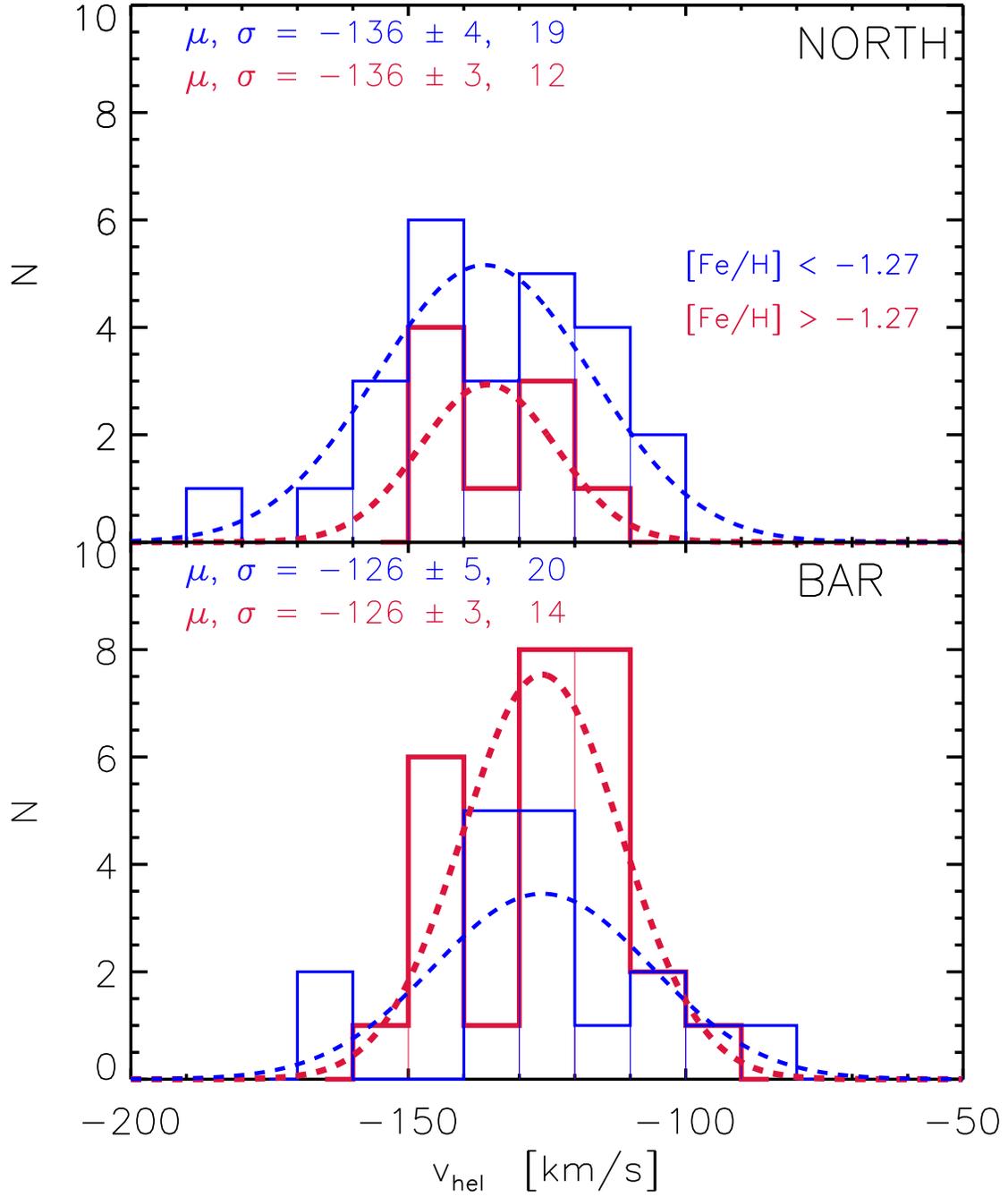}
\caption{ Velocity distributions for 
metallicity segregated subsamples in the north field (\textit{top panel}), 
and the bar field (\textit{bottom panel}).  
The average metallicity and dispersion values from the Gaussian fits 
(\textit{dashed lines}) are shown for all stars in each field near the 
top of each panel.  The mean uncertainty for each heliocentric corrected 
radial velocity value was $\pm 6$ km s$^{-1}$.}
\label{fig:2fvdf}
\end{center}
\end{figure}

\clearpage

\begin{figure}
\begin{center}
\plotone{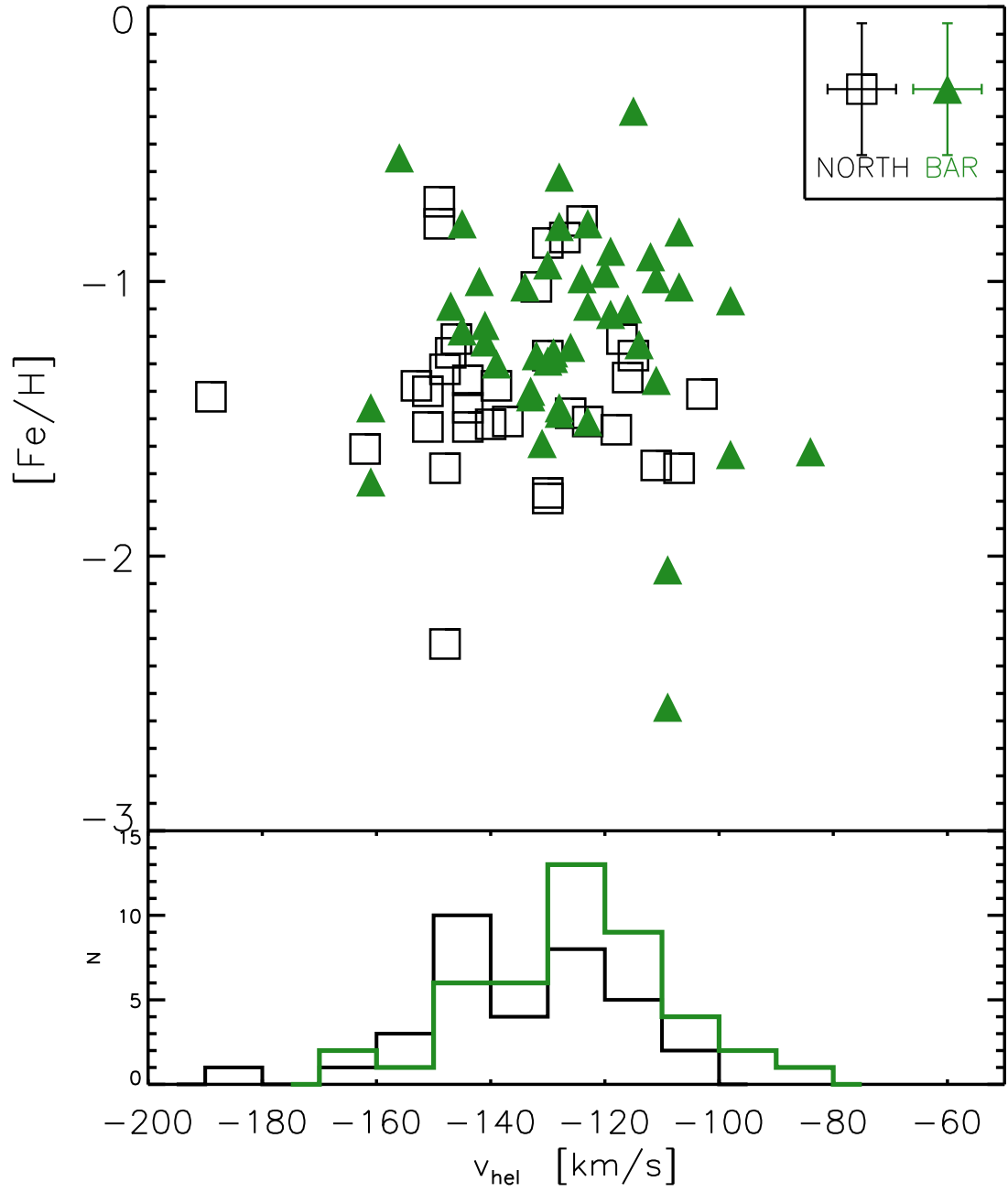}
\caption{Plot of the heliocentric velocity and metallicity of the 
78 RGB stars in our two WLM fields.  
The mean bar(\textit{green triangles}) and north (\textit{open squares}) field values 
are noticeably offset in both chemical and velocity space. The velocity offset between the two fields is shown more clearly by the histogram in the subpanel.  Errors are $\pm 0.25$ dex, 
and $\pm 6$ km s$^{-1}$ respectively per star, 
and shown for comparison on the simulated legend points at the top.}
\label{fig:vvmet}
\end{center}
\end{figure}

\clearpage

\begin{figure}
\begin{center}
\includegraphics[angle=90,width=1.0\textwidth]{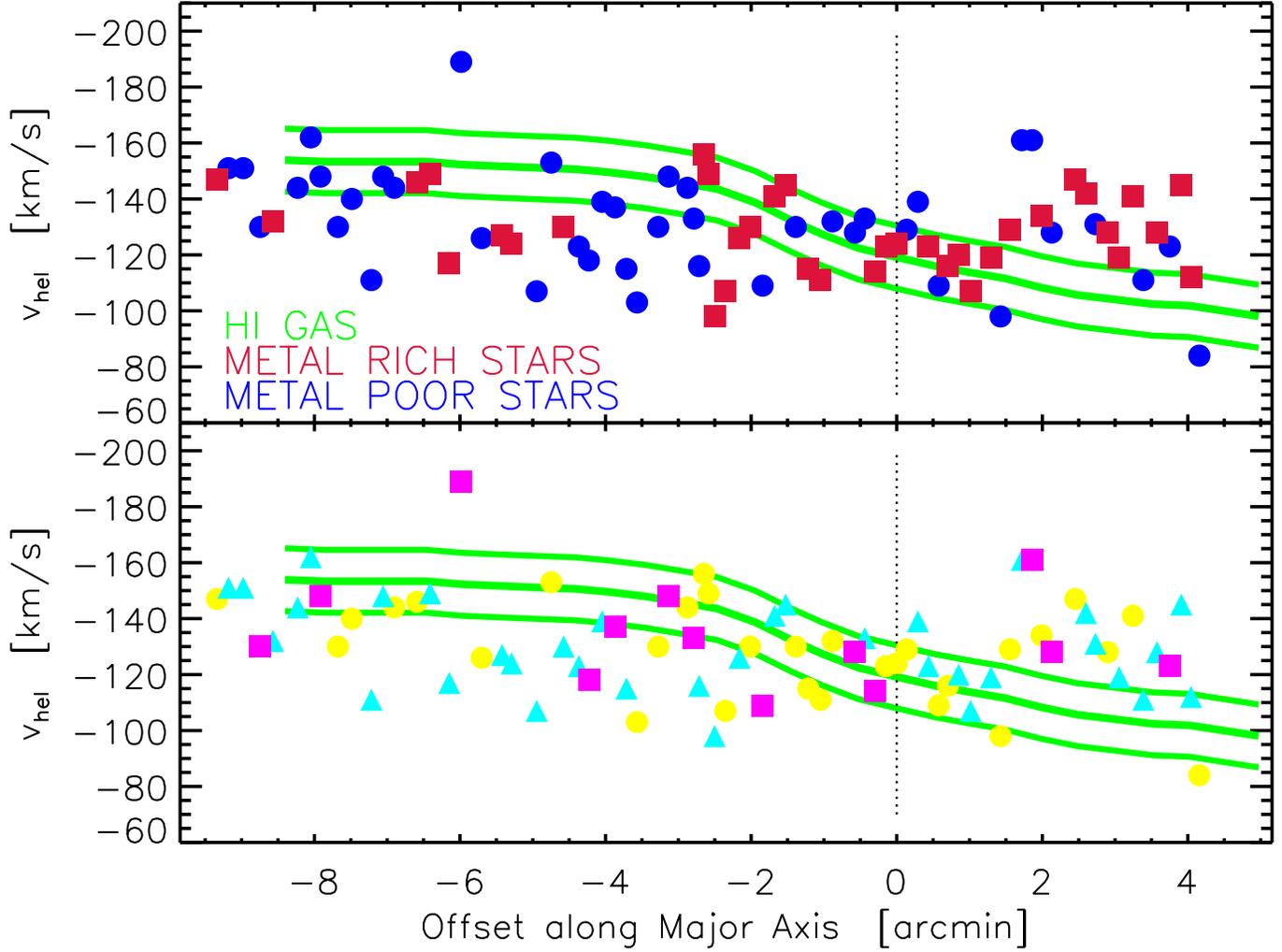}
\caption{(\emph{top panel}) Comparison of the heliocentric stellar velocity data for the 
metal rich (red), metal poor (blue) WLM subpopulations.  Shown in green is the \cite{Kepley07} HI gas position-velocity curve. 
Neither subpopulation tracks the gas particularly well, and the velocity gradient is only half that seen for the HI data (see text).  (\emph{bottom panel})  Similar visualization, but with the WLM stars subdivided into three age bins: $t\leq 1$ Gyr (\emph{circles}), $1\leq t \leq 6$Gyr (\emph{triangles}), $t\geq 6$ Gyr (\emph{squares}).  Negative offsets along the major axis correspond to the north portion of the galaxy, postive offsets the southern direction.}
\label{fig:pvnp}
\end{center}
\end{figure}

\clearpage

\begin{figure}
\begin{center}
\includegraphics[angle=90,width=1.0\textwidth]{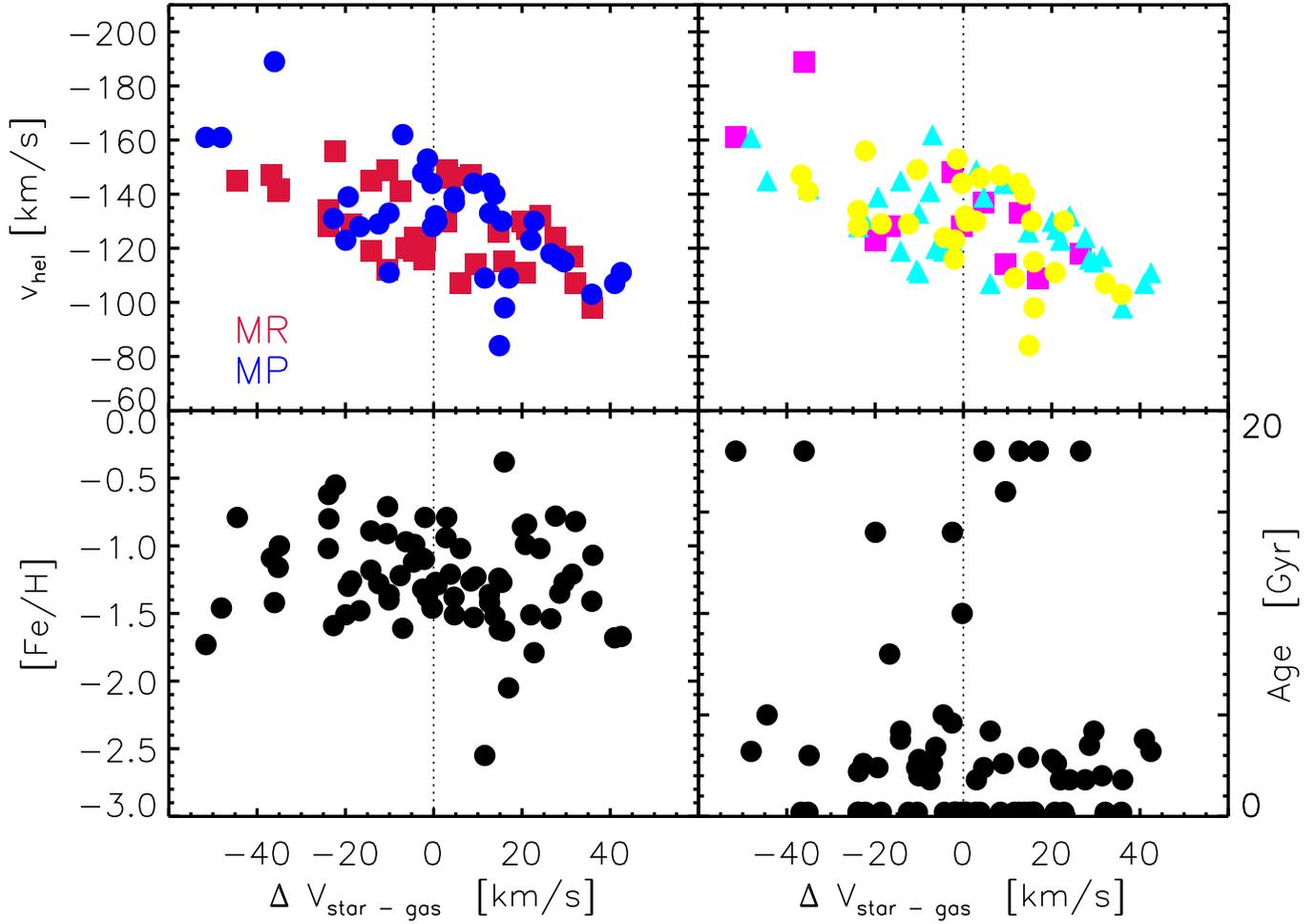}
\caption{Shown is the stellar to gas velocity offset (see text) for all stars in our sample, compared against various parameters.  Upper panels show the $\Delta V_{(star-gas)}$ values versus $v_{hel}$ for metallicity bins, and age bins (symbols are the same as in Figure \ref{fig:pvnp}).  (\emph{bottom panels}) $\Delta V_{(star-gas)}$ plotted versus [Fe/H] and age.  No clear trend exists between the star-gas velocity offset and the metallicity or age values.}
\label{fig:gdelt}
\end{center}
\end{figure}

\clearpage

\begin{deluxetable}{ccc}
\tablecolumns{3}
\tablewidth{0pc}
\tablecaption{WLM Properties}
\tablehead{
\colhead{Quantity} & \colhead{Value} & \colhead{Reference}
}
\startdata
$(l, b)$ & (75.85, -73.63) & \cite{Gallouet75}\\
E(B-V) & $0.035$ (mag.) & \cite{Alan05}\\
Distance & $932 \pm 33$ kpc & \cite{Alan05}\\
Eccentricity & 0.59 & \cite{Ables77}\\
Position Angle & 181 (deg.) & \cite{Jackson04}\\
Heliocentric Velocity ($v_{hel}^{HI}$) & $-130$ km s$^{-1}$ & \cite{Jackson04}\\
Rotation Velocity ($v_{rot}^{HI}$) & 30 km s$^{-1}$ & \cite{Kepley07}\\
M$_{\rm v}$ & 14.1 (mag.) & \cite{vandenBergh94}\\
$M_{dyn}$ & $2.16 \times 10^{9} M_{\odot}$ & \cite{Kepley07}\\
$M_{HI}$ & $(6.3 \pm 0.3) \times 10^{7} M_{\odot}$ & \cite{Kepley07}\\
$[Fe/H]^{RGB}_{phot}$ & $-1.45 \pm 0.2$ dex & \cite{Minniti97}\\
\enddata
\end{deluxetable}
\clearpage

\begin{deluxetable}{ccccccc} 
\tablecolumns{7} 
\tablewidth{0pc} 
\tablecaption{Observational Parameters} 
\tablehead{
\colhead{Target} & \colhead{Mag. Range} & \colhead{Distance (kpc)} & \colhead{R.A.} & \colhead{Dec.} & \colhead{Exp. Time (minutes)}
}
\startdata
WLM-N    & $22.1 \leq V_{mag} \leq 24.0$ & 970\tablenotemark{a} & 00:01:58 & -15:21:47 & 320\\
WLM-B    & $22.2 \leq V_{mag} \leq 23.2$ & 970\tablenotemark{a} & 00:01:58 & -15:28:30 & 320\\
NGC 104  & $11 \leq V_{mag} \leq 14$ & 4.5 & 00:24:05 & -72:04:51 & 1\\
NGC 1851 & $12 \leq V_{mag} \leq 16$ & 12.1 & 05:14:06 & -40:02:50 & 1\\
NGC 1904 & $12 \leq V_{mag} \leq 16$ & 12.9 & 05:24:10 & -24:31:27 & 1\\
NGC 7078 & $12 \leq V_{mag} \leq 16$ & 10.3 & 21:29:58 & +12:10:01 & 1\\
\enddata
\tablecomments{The location and distance data for the four calibrating globular clusters is taken from the \cite{Harris96} catalogue which can be found online at http://physwww.mcmaster.ca/$\sim$harris/mwgc.dat}
\tablenotetext{a}{From \cite{Gieren08}}
\end{deluxetable}
\clearpage

\begin{deluxetable}{cccccccccccccc}
\tablecolumns{14}
\tablewidth{0pc}
\tabletypesize{\small}
\tablecaption{Selected Parameters For WLM Stellar Sample}
\tablehead{
\colhead{STARID} & \colhead{R.A.} & \colhead{Dec.} &\colhead{r$_{ell}$} &\colhead{V} &\colhead{I} & \colhead{$W_{1}$} & \colhead{$W_{2}$} & \colhead{$W_{3}$} & \colhead{[Fe/H]$_{CaT}$} & \colhead{$\Delta$[Fe/H]} & \colhead{V$_{hel}$} & \colhead{$\Delta$V$_{hel}$} & \colhead{Age}\tablenotemark{a}\\
\colhead{} & \colhead{(degrees)} & \colhead{(degrees)} & \colhead{(degrees)} & \colhead{(mag.)} & \colhead{(mag.)} & \colhead{(${\rm \AA}$)} & \colhead{(${\rm \AA}$)} & \colhead{(${\rm \AA}$)} & \colhead{(dex)} & \colhead{(dex)} & \colhead{(km s$^{-1}$)} & \colhead{(km s$^{-1}$)} & \colhead{(Gyr)}
}
\startdata
27626 & 0.5293 & -15.5334 & 0.0978074 & 22.54 & 21.43 & 1.076 & 2.108 & 2.781 & -1.62 & 0.21 & -84 & 7 & $\leq 1.6$ \\
28986 & 0.4753 & -15.5315 & 0.0889742 & 22.82 & 21.36 & 1.209 & 3.408 & 3.208 & -0.91 & 0.26 & -112 & 8 & 2.4 \\
29674 & 0.4773 & -15.5292 & 0.0857606 & 22.96 & 21.25 & 1.250 & 3.316 & 3.527 & -0.79 & 0.27 & -145 & 9 & 5.0 \\
28080 & 0.4978 & -15.5266 & 0.0814584 & 22.64 & 21.09 & 1.072 & 2.972 & 2.158 & -1.51 & 0.22 & -123 & 6 & 14.0 \\
28895 & 0.4971 & -15.5237 & 0.0775812 & 22.80 & 21.35 & 1.551 & 3.819 & 2.791 & -0.80 & 0.27 & -128 & 7 & 2.2 \\
27491 & 0.5058 & -15.5205 & 0.0747572 & 22.51 & 21.19 & 1.255 & 2.680 & 2.821 & -1.36 & 0.23 & -111 & 7 & 2.0 \\
26541 & 0.4892 & -15.5182 & 0.0702356 & 22.25 & 20.97 & 1.604 & 3.500 & 2.439 & -1.16 & 0.25 & -141 & 7 & $\leq 1.6$ \\
27842 & 0.4915 & -15.5150 & 0.0661231 & 22.59 & 20.35 & 1.332 & 3.647 & 3.089 & -0.89 & 0.27 & -119 & 7 & 3.8 \\
27409 & 0.4978 & -15.5124 & 0.0630628 & 22.49 & 21.00 & 1.643 & 3.800 & 3.509 & -0.62 & 0.28 & -128 & 7 & $\leq 1.6$ \\
27569 & 0.4726 & -15.5096 & 0.0620014 & 22.53 & 21.22 & 1.032 & 2.836 & 2.177 & -1.59 & 0.21 & -131 & 6 & 2.6 \\
27916 & 0.4639 & -15.5074 & 0.0626483 & 22.61 & 20.96 & 1.735 & 3.463 & 2.533 & -1.00 & 0.25 & -142 & 7 & 3.0 \\
27018 & 0.4707 & -15.5049 & 0.0569988 & 22.39 & 21.12 & 1.284 & 3.317 & 3.305 & -1.09 & 0.25 & -147 & 7 & $\leq 1.6$ \\
27845 & 0.4883 & -15.4996 & 0.0461610 & 22.59 & 21.10 & 1.351 & 3.373 & 1.615 & -1.48 & 0.22 & -128 & 6 & 8.0 \\
28137 & 0.5336 & -15.4972 & 0.0606404 & 22.65 & 21.39 & 1.065 & 4.178 & 2.393 & -1.02 & 0.26 & -134 & 8 & $\leq 1.6$ \\
28441 & 0.5078 & -15.4951 & 0.0435499 & 22.71 & 21.12 & 0.860 & 2.543 & 2.087 & -1.73 & 0.20 & -161 & 9 & 18.0 \\
28427 & 0.4690 & -15.4927 & 0.0435044 & 22.71 & 21.37 & 1.318 & 3.107 & 1.863 & -1.46 & 0.22 & -161 & 6 & 3.2 \\
27310 & 0.4617 & -15.4900 & 0.0449992 & 22.46 & 21.29 & 1.502 & 3.167 & 2.398 & -1.26 & 0.24 & -129 & 7 & $\leq 1.6$ \\
26940 & 0.4765 & -15.4879 & 0.0342952 & 22.37 & 21.13 & 1.307 & 3.107 & 1.655 & -1.63 & 0.22 &  -98 & 7 & $\leq 1.6$ \\
28333 & 0.4608 & -15.4856 & 0.0416024 & 22.69 & 21.14 & 0.996 & 3.479 & 2.841 & -1.12 & 0.25 & -119 & 6 & 5.0 \\
28970 & 0.5081 & -15.4810 & 0.0277182 & 22.82 & 21.25 & 1.38 & 3.03 & 3.097 & -1.02 & 0.25 & -107 & 6 & 4.2 \\
28881 & 0.5140 & -15.4783 & 0.0293921 & 22.80 & 21.31 & 1.269 & 3.353 & 3.056 & -0.97 & 0.26 & -120 & 7 & 3.4 \\
27265 & 0.5193 & -15.4758 & 0.0320861 & 22.45 & 21.34 & 1.295 & 3.567 & 2.699 & -1.10 & 0.25 & -116 & 6 & $\leq 1.6$ \\
27427 & 0.4687 & -15.4737 & 0.0260121 & 22.49 & 21.75 & 0.971 & 1.304 & 0.962 & -2.55 & 0.13 & -109 & 9 & $\leq 1.6$ \\
27926 & 0.4824 & -15.4713 & 0.0129957 & 22.61 & 21.00 & 1.505 & 3.346 & 2.603 & -1.09 & 0.25 & -123 & 6 & 4.6 \\
28395 & 0.4847 & -15.4689 & 0.0091797 & 22.70 & 21.41 & 1.213 & 3.563 & 1.996 & -1.30 & 0.24 & -139 & 6 & 2.4 \\
26678 & 0.5100 & -15.4664 & 0.0190786 & 22.29 & 21.40 & 0.977 & 3.412 & 2.772 & -1.28 & 0.25 & -129 & 5 & $\leq 1.6$ \\
27833 & 0.5018 & -15.4641 & 0.0106341 & 22.59 & 21.68 & 1.572 & 3.131 & 3.069 & -0.99 & 0.25 & -124 & 6 & $\leq 1.6$ \\
27429 & 0.5069 & -15.4616 & 0.0160755 & 22.49 & 21.13 & 1.703 & 3.674 & 3.058 & -0.79 & 0.27 & -123 & 7 & $\leq 1.6$ \\
28322 & 0.4653 & -15.4592 & 0.0269871 & 22.69 & 20.90 & 1.640 & 2.952 & 2.410 & -1.23 & 0.23 & -114 & 6 & 16.0 \\
27407 & 0.5182 & -15.4567 & 0.0287657 & 22.49 & 21.11 & 1.358 & 3.259 & 2.033 & -1.40 & 0.23 & -133 & 7 & 2.8 \\
28405 & 0.5148 & -15.4545 & 0.0267520 & 22.70 & 21.18 & 1.132 & 2.586 & 2.590 & -1.46 & 0.22 & -128 & 6 & 10.0 \\
27044 & 0.5086 & -15.4494 & 0.0258708 & 22.40 & 21.24 & 1.241 & 3.385 & 2.467 & -1.27 & 0.24 & -132 & 6 & $\leq 1.6$ \\
28316 & 0.5162 & -15.4466 & 0.0338115 & 22.69 & 21.28 & 1.241 & 3.385 & 2.467 & -0.99 & 0.25 & -111 & 6 & $\leq 1.6$ \\
26856 & 0.4692 & -15.4438 & 0.0344805 & 22.35 & 20.94 & 1.991 & 4.473 & 3.291 & -0.38 & 0.31 & -115 & 6 & $\leq 1.6$ \\
27091 & 0.4726 & -15.4410 & 0.0354185 & 22.41 & 21.25 & 1.273 & 3.422 & 2.340 & -1.29 & 0.24 & -130 & 6 & $\leq 1.6$ \\
28220 & 0.4741 & -15.4387 & 0.0372675 & 22.67 & 21.17 & 0.903 & 3.485 & 2.757 & -1.18 & 0.25 & -145 & 6 & 4.2 \\
27342 & 0.5131 & -15.4361 & 0.0424025 & 22.47 & 21.09 & 1.040 & 3.951 & 2.198 & -1.22 & 0.25 & -141 & 5 & 1.8 \\
28371 & 0.5223 & -15.4334 & 0.0506345 & 22.70 & 21.31 & 1.122 & 2.208 & 1.219 & -2.05 & 0.17 & -109 & 6 & 18.0 \\
28056 & 0.5086 & -15.4305 & 0.0469298 & 22.63 & 21.28 & 1.263 & 3.507 & 3.106 & -0.94 & 0.26 & -130 & 6 & $\leq 1.6$ \\
27590 & 0.5008 & -15.4281 & 0.0476470 & 22.53 & 21.16 & 1.048 & 3.247 & 2.778 & -1.24 & 0.24 & -126 & 7 & 2.9 \\
27004 & 0.5074 & -15.4248 & 0.0534145 & 22.39 & 21.22 & 1.518 & 3.831 & 3.092 & -0.82 & 0.27 & -107 & 9 & $\leq 1.6$ \\
27856 & 0.5229 & -15.4224 & 0.0627850 & 22.59 & 21.13 & 1.721 & 3.454 & 2.367 & -1.07 & 0.25 &  -98 & 8 & 1.8 \\
27548 & 0.4668 & -15.4210 & 0.0611249 & 22.52 & 21.20 & 1.716 & 3.654 & 3.285 & -0.71 & 0.28 & -149 & 7 & $\leq 1.6$ \\
28328 & 0.5189 & -15.4199 & 0.0637268 & 22.69 & 21.06 & 1.682 & 4.416 & 2.915 & -0.55 & 0.29 & -156 & 6 & $\leq 1.6$ \\
28682 & 0.4792 & -15.4188 & 0.0599319 & 22.76 & 21.43 & 0.787 & 3.122 & 2.674 & -1.35 & 0.23 & -116 & 8 & 3.5 \\
31087 & 0.4885 & -15.4176 & 0.0603162 & 23.21 & 21.47 & 0.996 & 2.930 & 2.110 & -1.42 & 0.21 & -133 & 6 & 18.0 \\
27765 & 0.4819 & -15.4161 & 0.0628594 & 22.57 & 21.41 & 1.439 & 2.975 & 1.982 & -1.46 & 0.22 & -144 & 6 & $\leq 1.6$ \\
28590 & 0.4775 & -15.4118 & 0.0691717 & 22.75 & 21.13 & 1.329 & 3.127 & 2.235 & -1.32 & 0.23 & -148 & 6 & 14.0 \\
27189 & 0.4786 & -15.4095 & 0.0719565 & 22.44 & 21.26 & 1.509 & 3.021 & 2.529 & -1.27 & 0.24 & -130 & 6 & $\leq 1.6$ \\
26226 & 0.5257 & -15.4046 & 0.0845370 & 22.13 & 20.84 & 1.456 & 2.996 & 2.437 & -1.41 & 0.23 & -103 & 6 & $\leq 1.6$ \\
28196 & 0.4656 & -15.4022 & 0.0843228 & 22.66 & 21.26 & 0.946 & 3.665 & 2.272 & -1.27 & 0.24 & -115 & 6 & 4.2 \\
28736 & 0.4630 & -15.3996 & 0.0883728 & 22.77 & 21.25 & 1.082 & 3.432 & 1.590 & -1.51 & 0.22 & -137 & 6 & 18.0 \\
28821 & 0.5198 & -15.3966 & 0.0920141 & 22.79 & 21.51 & 1.067 & 3.489 & 1.903 & -1.38 & 0.23 & -139 & 8 & 2.4 \\
29499 & 0.5074 & -15.3936 & 0.0928048 & 22.92 & 21.41 & 1.078 & 2.906 & 1.921 & -1.54 & 0.21 & -118 & 6 & 18.0 \\
27734 & 0.5055 & -15.3913 & 0.0954310 & 22.57 & 21.28 & 1.539 & 2.910 & 1.823 & -1.51 & 0.21 & -123 & 7 & 1.8 \\
29590 & 0.4825 & -15.3878 & 0.0992865 & 22.94 & 21.49 & 1.346 & 4.305 & 2.224 & -0.86 & 0.27 & -130 & 9 & 2.8 \\
27667 & 0.4857 & -15.3850 & 0.1026929 & 22.55 & 21.30 & 0.789 & 3.592 & 2.271 & -1.38 & 0.24 & -153 & 6 & $\leq 1.6$ \\
27192 & 0.4952 & -15.3816 & 0.1069359 & 22.44 & 21.00 & 1.148 & 2.765 & 1.947 & -1.68 & 0.21 & -107 & 8 & 3.8 \\
27515 & 0.5155 & -15.3759 & 0.1168678 & 22.51 & 20.92 & 1.806 & 3.416 & 3.241 & -0.78 & 0.27 & -124 & 6 & 1.8 \\
28531 & 0.4731 & -15.3737 & 0.1185451 & 22.73 & 21.15 & 1.946 & 3.563 & 2.593 & -0.84 & 0.26 & -127 & 6 & 2.6 \\
28064 & 0.5206 & -15.3691 & 0.1266431 & 22.64 & 21.42 & 1.198 & 2.689 & 2.413 & -1.48 & 0.22 & -126 & 6 & $\leq 1.6$ \\
19203 & 0.5036 & -15.3643 & 0.1298484 & 24.05 & 21.47 & 0.970 & 2.766 & 1.639 & -1.42 & 0.19 & -189 & 7 & 18.0 \\
16415 & 0.5188 & -15.3616 & 0.1356525 & 22.59 & 21.21 & 1.645 & 3.217 & 2.266 & -1.21 & 0.24 & -117 & 5 & 2.0 \\
16348 & 0.4626 & -15.3572 & 0.1414459 & 22.52 & 21.06 & 1.378 & 4.115 & 2.913 & -0.79 & 0.28 & -149 & 6 & 1.8 \\
16163 & 0.5051 & -15.3542 & 0.1430462 & 22.30 & 21.10 & 1.103 & 3.501 & 2.765 & -1.21 & 0.25 & -146 & 6 & $\leq 1.6$ \\
16218 & 0.5013 & -15.3490 & 0.1494224 & 22.38 & 21.11 & 1.325 & 3.064 & 2.466 & -1.36 & 0.23 & -144 & 6 & $\leq 1.6$ \\
16335 & 0.5224 & -15.3465 & 0.1555663 & 22.51 & 21.22 & 0.667 & 2.320 & 0.911 & -2.32 & 0.16 & -148 & 6 & 3.2 \\
16326 & 0.5105 & -15.3438 & 0.1570543 & 22.50 & 21.12 & 1.017 & 2.970 & 1.857 & -1.67 & 0.21 & -111 & 7 & 3.2 \\
16329 & 0.4811 & -15.3393 & 0.1619626 & 22.50 & 21.30 & 1.208 & 2.929 & 2.136 & -1.52 & 0.22 & -140 & 6 & $\leq 1.6$ \\
16225 & 0.4775 & -15.3361 & 0.1663648 & 22.39 & 21.21 & 0.629 & 3.286 & 1.653 & -1.79 & 0.21 & -130 & 5 & $\leq 1.6$ \\
16511 & 0.5299 & -15.3321 & 0.1753224 & 22.69 & 21.23 & 0.760 & 3.408 & 1.502 & -1.68 & 0.21 & -148 & 6 & 8.0 \\
16354 & 0.4947 & -15.3299 & 0.1738444 & 22.53 & 21.21 & 0.797 & 3.207 & 1.974 & -1.61 & 0.21 & -162 & 6 & 2.6 \\
16298 & 0.4728 & -15.3269 & 0.1786507 & 22.46 & 21.09 & 1.175 & 3.025 & 2.082 & -1.53 & 0.22 & -144 & 6 & 2.6 \\
16413 & 0.4822 & -15.3213 & 0.1852281 & 22.59 & 21.13 & 1.197 & 3.741 & 2.747 & -1.02 & 0.26 & -132 & 7 & 1.8 \\
16474 & 0.5200 & -15.3183 & 0.1910044 & 22.65 & 21.11 & 0.867 & 2.697 & 1.853 & -1.77 & 0.20 & -130 & 5 & 16.0 \\
16299 & 0.5024 & -15.3144 & 0.1941447 & 22.46 & 21.07 & 0.731 & 3.259 & 2.277 & -1.53 & 0.23 & -151 & 5 & 3.2 \\
16259 & 0.4855 & -15.3111 & 0.1982649 & 22.43 & 21.04 & 0.937 & 3.042 & 2.702 & -1.40 & 0.23 & -151 & 6 & 2.7 \\
16300 & 0.5033 & -15.3084 & 0.2019618 & 22.46 & 21.20 & 1.605 & 2.852 & 2.630 & -1.26 & 0.24 & -147 & 9 & $\leq 1.6$ \\
\enddata
\tablenotetext{a}{Derived ages were cutoff artificially at 1.6 Gyr as the interpolation of the model space became uncertain below this age.  Stars following below this were assigned an age of 0.2 Gyr to aid in interpretation of the sample.}
\end{deluxetable}
\clearpage

\begin{deluxetable}{lccc} 
\tablecolumns{4} 
\tablewidth{0pc} 
\tablecaption{Binned Age Metallicity Statistics} 
\tablehead{
\colhead{Variable} & \colhead{$ t < 1$} & \colhead{$ 1 \leq t \leq 6 $} & \colhead{$ t > 6 $}
}
\startdata
$\langle [Fe/H] \rangle$ (dex) & $-1.21 \pm 0.05$ & $-1.23 \pm 0.04$ & $-1.55 \pm 0.07$\\
$\langle \sigma_{[Fe/H]} \rangle$ (dex) & 0.42 & 0.34 & 0.22\\
$\langle v_{hel} \rangle$ (km s$^{-1}$) & $-129 \pm 1$ & $-130 \pm 1$ & $-136 \pm 2$\\
$\langle \sigma_{v} \rangle$ (km s$^{-1})$ & 17 & 16 & 22\\
$\langle t \rangle$ (Gyr) & 0.2 & 3.0 & 15\\
$\langle r_{ell} \rangle$ (deg.) & 0.07 & 0.09 & 0.08\\
$\langle \sigma_{r} \rangle$ (deg.) & 0.05 & 0.06 & 0.05\\
      $n$ & 30 & 35 & 13\\
\enddata
\tablecomments{Average values in three age bins, for various parameters of our WLM stellar sample.}
\end{deluxetable}
\clearpage

\begin{deluxetable}{cccc}
\tablecolumns{4}
\tablewidth{0pc}
\tablecaption{Subpopulation $\frac{v_{rot}}{\sigma_{v}}$ Ratios}
\tablehead{
\colhead{Component} & \colhead{$v_{rot}$} &\colhead{$\sigma_{v}$} & \colhead{$\frac{v_{rot}}{\sigma_{v}}$}
}
\startdata
Full Sample & 20.78 & 17.53 & 1.19\\
Metal Rich & 20.61 & 14.29 & 1.44\\
Metal Poor & 19.64 & 19.91 & 0.99\\
$t < 1$ Gyr & 22.55 & 17.31 & 1.30\\
$1 \leq t \leq 6$ Gyr & 17.52 & 16.14 & 1.09\\
$t > 6$ Gyr & 9.90 & 15.20 & 0.65\\
HI-like velocities ("disk") & 40.50 & 13.50 & 3.00\\
non-HI-like velocities ("halo") & 22.86 & 18.67 & 1.23\\
\enddata
\tablecomments{The final two rows separate the sample based on those stars that fall in the same position-velocity space as the \cite{Kepley07} HI velocity data.}
\end{deluxetable}
\clearpage

\end{document}